\documentclass[11pt,a4paper]{article}
\usepackage[T1]{fontenc}
\usepackage[latin1,utf8]{inputenc}
\usepackage{cancel}
\usepackage{amsmath,amssymb,amsfonts,multicol}
\usepackage[normal,font=small,labelfont=bf,labelsep=period]{caption}
\usepackage[pdftex]{color,graphicx}
\usepackage[english]{babel}
\usepackage[compress]{cite}
\usepackage[dvipsnames]{xcolor}
\usepackage{slashed}
\usepackage{enumitem}
\usepackage{cite}
\usepackage{mathtools}
\usepackage[colorlinks=false,linkcolor=blue,urlcolor=magenta,citecolor=blue]{hyperref}
\hypersetup{%
 colorlinks = true,
linkcolor  = black}

\newcommand{\be}{\begin{equation}}
\newcommand{\ee}{\end{equation}}
\newcommand{\bea}{\begin{eqnarray}}
\newcommand{\eea}{\end{eqnarray}}
\newcommand{\eq}[1]{(\ref{#1})}

\numberwithin{equation}{section}

\definecolor{colorn}{rgb}{.1,.1,.8}

\begin{document}

\pagenumbering{gobble}

\begin{flushright}
\footnotesize
{IFT-UAM/CSIC-19-163}
\end{flushright}
\vspace{1cm}

\begin{center}

{\LARGE\color{black}\bf Detuning primordial black hole dark matter\\  with early matter domination and axion monodromy \\[1mm] }

\medskip
\bigskip\color{black}\vspace{1cm}
{
{\bf Guillermo Ballesteros}\,$^{1,2}$, {\bf Julián Rey}\,$^{1,2}$ and {\bf{Fabrizio Rompineve}\,$^{3,4}$}
}
\\[7mm]
\small \it{$^{1}$Instituto de F\'isica Te\'orica UAM/CSIC, Calle Nicolás Cabrera 13--15\\ Cantoblanco E-28049 Madrid, Spain  \\\vspace{0.1cm}
$^2$Departamento de F\'isica Te\'orica, Universidad Aut\'onoma de Madrid (UAM)\\ Campus de Cantoblanco, E-28049 Madrid, Spain\\\vspace{0.1cm}
$^{3}$ Institute of Cosmology, Dept. of Physics and Astronomy, Tufts University\\
Medford, MA 02155, USA\\
\vspace{0.1cm}
$^{4}$ IFAE and BIST, Campus UAB, E-08193 Bellaterra, Barcelona, Spain}
\end{center}
\vspace{1cm}

\begin{center}\textbf{Abstract}\end{center}

We present a scenario that ameliorates the tuning problems present in models of primordial black hole (PBH) dark matter from inflation. Our setup employs the advantages of gravitational collapse in a long epoch of early matter domination with reheating temperature $\lesssim 10^{6}~\text{GeV}$. Furthermore, we make use of a string-inspired class of  models where the inflaton is identified with a non-compact axion field. In this framework, the presence of multiple local minima in the inflaton potential can be traced back to an approximate discrete shift symmetry. This scenario allows the formation of {PBHs} in the observationally viable range of masses ($M_{\text{PBH}}\sim 10^{-16}M_{\odot}-10^{-13}M_{\odot}$) accounting for all dark matter,  and in excellent agreement with the CMB. We find a significant reduction in the required tuning of the parameters of the inflationary potential, in contrast to the standard case of {PBH} formation during radiation domination. {However, abundant formation of light PBHs during an early phase of matter domination can be more easily in conflict with evaporation bounds. We discuss how these can be avoided under mild assumptions on the collapsing energy density fraction.}

\begin{center}
\vfill\flushleft
\noindent\rule{6cm}{0.4pt}\\
{\small  \tt guillermo.ballesteros@uam.es, julian.rey@uam.es, fabrizio.rompineve@tufts.edu}
\end{center}

\newpage
\pagenumbering{arabic}
\vspace{1.5cm}
\tableofcontents

\section{Introduction}
\label{intro}

Primordial black holes {(PBHs)~\cite{Zeldovich, Carr:1974nx,1979A&A....80..104N}} are intriguing candidates for the dark matter (DM) of the Universe~\cite{Chapline}. On the theoretical side, their formation may be a subproduct of inflation, so they are often regarded as a more economical explanation of DM than particle physics proposals such as axions or weakly interacting massive particles (WIMPs). On the observational side, PBHs are macroscopic objects and thus exhibit a variety of astrophysical signatures; see \cite{Sasaki:2018dmp} for a non exhaustive list. At the time of writing, PBHs can account  for all  the DM provided that their mass is in the window\footnote{See~\cite{Katz:2018zrn,Montero-Camacho:2019jte} for critical takes on previously claimed constraints in this range.}
\begin{align} \label{win}
10^{-16}~M_{\odot}\lesssim M_{\text{PBH}} \lesssim 10^{-11}M_{\odot}\,.
\end{align} 
The lower end of this mass range comes from extragalactic gamma-ray observations \cite{Carr:2009jm,Arbey:2019vqx,Ballesteros:2019exr} and the  upper one is due to microlensing \cite{Niikura:2017zjd}. {Further bounds related to the possible evaporation of PBHs in the Galaxy exist --with diverse assumptions concerning the mass distribution and spin-- at the lower end of the range: $e^\pm$ measurements by the Voyager probe \cite{Boudaud:2018hqb}, positron-electron annihilation into 511 keV photon emission susceptible of being detected by the telescope INTEGRAL \cite{DeRocco:2019fjq,Laha:2019ssq,Dasgupta:2019cae}, the  emission of a neutrino flux that could be detected by Superkamiokande \cite{Dasgupta:2019cae}, modifications on the CMB anisotropies due to energy injection  around  the time of recombination \cite{Poulin:2016anj,Stocker:2018avm}, and the Galactic emission of gamma/X-rays as measured by the satellites Fermi \cite{Carr:2016hva} and INTEGRAL \cite{Laha:2020ivk}. Among these, the strongest bound is the latest, which excludes non-spinning PBHs of mass $10^{17}$~g from being more than $\sim 10\%$ of the DM. We will mostly refer to extragalactic gamma-ray bounds (which do not suffer from uncertainties related to gamma-ray propagation) and mention other bounds wherever it is appropriate. 

Concrete realizations of the appealing idea of PBH DM} come with undeniable downsides. The most studied mechanism of PBH formation relies on the gravitational collapse of Hubble-sized regions triggered by large density fluctuations (seeded by inflation) upon horizon re-entry \cite{Carr:1975qj, Ivanov:1994pa}.\footnote{Non-inflationary mechanisms of PBH formation exist, but they generically require extra ingredients that reduce the original appeal of PBHs as DM. See~e.g.\cite{Vilenkin:1981iu, Fort:1993zb, Khlopov:2004sc, Deng:2016vzb, Ferrer:2018uiu} for PBH formation from the collapse of topological defects.} For PBHs to form abundantly --and contribute significantly to the DM--  these fluctuations, which are huge by CMB standards, must occur at distance scales that are about fifteen orders of magnitude smaller than the CMB ones. In the context of canonical single-field inflation, such a spectral feature can be obtained if the inflaton traverses a region of its potential, $V$, with an inflection point ($V''=0$), {see \cite{Starobinsky:1992ts} (and \cite{Ivanov:1994pa} for an early application in the context of PBH formation).} This can be either a local flattening of the potential or an inflection point between a local shallow minimum followed by a maximum. Either way, the inflaton loses kinetic energy in that region, which results in an enhancement of the primordial spectrum. Several models implementing this idea have been put forward in the last couple of years, see \cite{Garcia-Bellido:2017mdw,Ezquiaga:2017fvi,Kannike:2017bxn, Ballesteros:2017fsr,Hertzberg:2017dkh,Cicoli:2018asa,Ozsoy:2018flq,Dalianis:2018frf}. In general, the inflection point has to be introduced in the potential in an ad hoc manner. If the original appeal of PBH DM (and of inflation!) is to be preserved, having a motivation for the existence of such a region independently of PBH formation would  be highly desirable.

In addition, the mass of PBHs formed during RD scales as $\sim \exp(- 2N)$, where $N$ is the number of e-folds of inflation.\footnote{In our convention, $N$ grows as inflation proceeds.} Therefore,  in order to produce PBHs in the window \eq{win}, the feature of the potential must be placed in a narrow region of $N$, of width $\Delta N\simeq 6$. In a generic single-field inflation model capable of PBH formation, this generically constrains its parameters beyond the requirements imposed by the CMB. There are however models, such as the ones proposed in \cite{Ballesteros:2017fsr}, for which the requirement of enough inflation tends to lead to the location of the feature in the right ballpark of $N$. 

More worrisome than the relatively narrow $\Delta N$ is a different property shared by these scenarios with an inflection point: as it was pointed out in \cite{Ballesteros:2017fsr}, the magnitude of the required enhancement of the fluctuations is highly sensitive to the parameters of the potential. Larger spectral enhancements require a higher level of tuning  of the parameters. If the PBHs are formed during the radiation-dominated (RD) epoch, the primordial power spectrum must increase by $\sim 10^{7}$ with respect to its amplitude at CMB scales to account for all DM. Generically, this implies a severe fine tuning of the parameters of the potential. In summary: not only an ad hoc feature altering the potential has to be placed in a rather precise location, but also the details of its shape need to be carefully crafted. At the core of this issue lies the fact that the PBH abundance depends exponentially on the variance of the radiation  density fluctuations and therefore on the primordial power spectrum. 

The answer to the question of whether or not this strong parameter sensitivity is an utterly abhorrent property is a matter of personal taste. What is clear is that PBH formation is non-generic in inflation. We stress that this parameter sensitivity is also present in multi-field and hybrid inflation models of PBH formation. Given that PBHs have risen very recently as a popular DM candidate, we think it is interesting to explore ideas that can make PBH production a less ad hoc occurrence within inflation. 

In this paper  we present a well-motivated scenario that alleviates the aforementioned downsides. The key aspects of our setup are an inflationary potential which naturally features the existence of several local minima as a consequence of an underlying approximate discrete shift symmetry, and a matter dominated (MD) epoch right after inflation, during which the likeliness of gravitational collapse is augmented \cite{Carr:1975qj,Khlopov:1980mg,Harada:2016mhb}. We find that these two ingredients can lead to the formation of PBHs accounting for all DM (in the adequate mass ballpark), and the fit of the inflationary model to the CMB measurements tuns out to be excellent. This is in contrast to most proposals featuring an approximate inflection point, for which the scalar spectral index, $n_s$, at $k=0.05$ Mpc$^{-1}$ tends to be lower than the one determined by the Planck collaboration.\footnote{See \cite{Ozsoy:2018flq} for an exception to that trend on $n_s$ in another setup inspired by AMI. See appendix \ref{cosmoparams} (of our work) for a summary of the relevant cosmological parameters from the latest CMB data.}

\begin{figure}[t]
\includegraphics[scale=0.33]{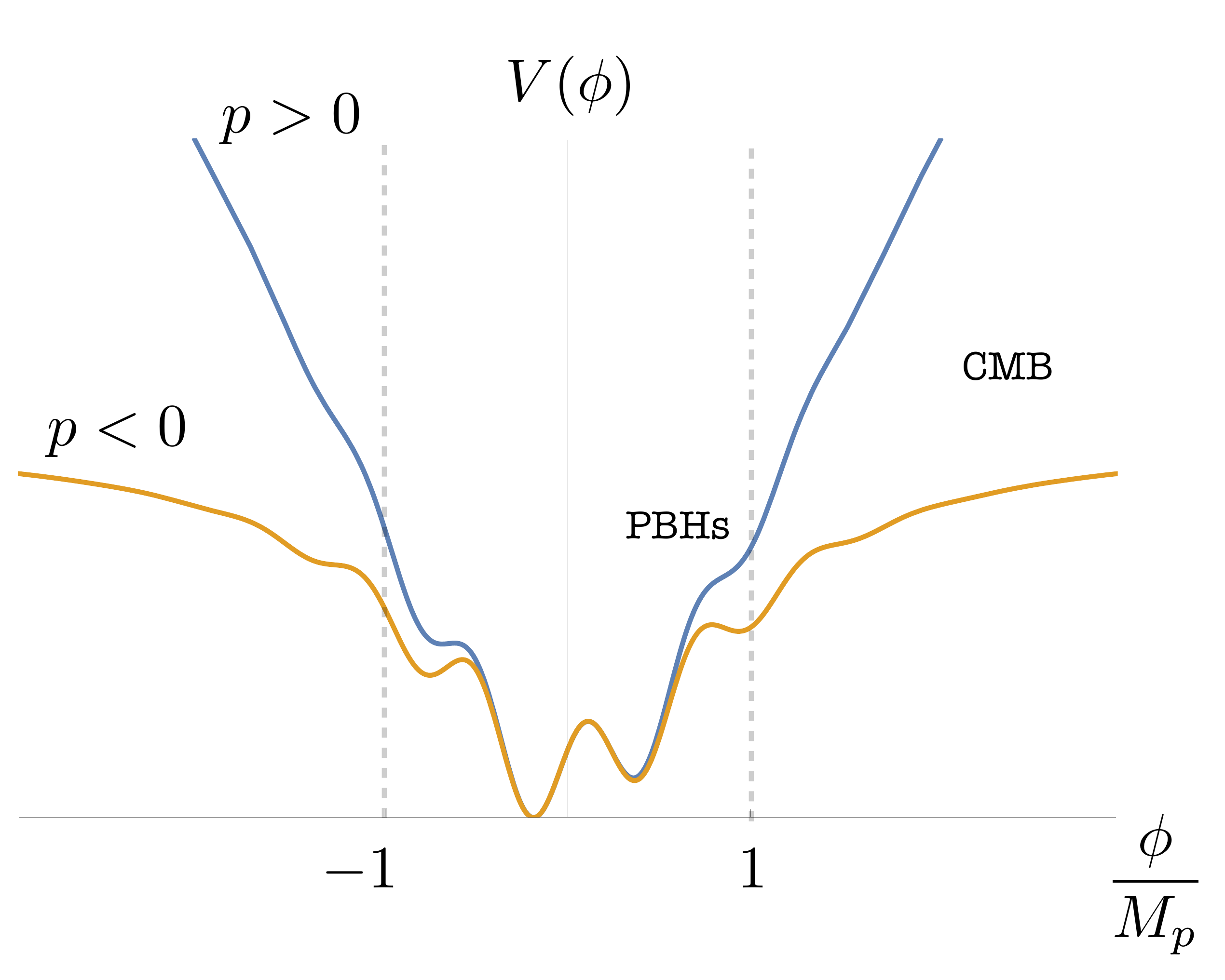} 
\centering
\caption{\it Inflationary potentials which will be considered in this work. Close to the global minimum, the potential exhibits {a modulation} superimposed on a quadratic potential. Inflationary fluctuations can then be enhanced by the presence of local minima and eventually lead to PBHs. At large superplanckian field values the potential flattens and grows as a power law for $p>0$, while it asymptotes to a plateau for $p<0$. In this region, slow-roll inflation can generate CMB anisotropies.}
\label{fig:potentials}
\end{figure}

Our scenario is inspired by axion monodromy inflation (AMI)~\cite{Silverstein:2008sg, McAllister:2008hb}. In this framework, the inflaton is a pseudo-scalar field with a discrete shift symmetry which is broken by a non-periodic potential term. The full inflaton potential features the characteristic axionic {modulation}, superimposed on the monodromic term. When the amplitude of this {modulation} is large enough, near-inflection points and local minima appear in the inflationary trajectory. 
AMI can arise from string compactifications, where the inflaton field is generically accompanied by other (heavier and initially stabilized) scalar fields, called moduli. As the inflaton travels $\gtrsim M_p$ distances during inflation, the heavy moduli tend to shift from their VEVs and backreact on the inflationary trajectory. This leads to flattening of the inflaton potential~\cite{Dong:2010in, McAllister:2014mpa} at large field values. The non-periodic part of the potential is generally quadratic for $\phi\lesssim M_{p}$, whereas it behaves as $V \sim b+\phi^{2p}$, with $p<1$ and $b$ constant, for $\phi\gtrsim M_{p}$. In certain realizations of AMI, which will be of particular interest for our work, the amplitude of the axionic potential {modulation} is suppressed at large field values~\cite{Flauger:2014ana}. Therefore, AMI can provide inflationary potentials which exhibit two distinct regions (see Figure~\ref{fig:potentials}): the first one, close to the global minimum, can feature the critical points that are desired for PBH formation; the second region, at large field values, does not display {modulation} and is instead ideal to realize large field inflation. In this work, we take a phenomenological approach to AMI and consider positive as well as negative rational values of $p$, while also requiring agreement with the latest CMB constraints~\cite{Akrami:2018odb}. Negative values of $p$ lead to plateau-like potential at large field values. This feature can indeed arise in AMI~\cite{Landete:2017amp}.

Our setup can naturally accommodate a long MD epoch after the end of inflation, before the inflaton decays completely and reheats the Universe. First of all, the minima of the potential for small field values can support small oscillations of the inflaton (which effectively take place  in a quadratic potential). If the oscillations of the inflaton in the minimum where inflation ends (assumed to be $V=0$) dominate the energy budget of the Universe, the latter enters into a phase of MD after {inflation; see \cite{1978PAZh....4..155S} (and also \cite{Turner:1983he} for a generalization to other equations of state).} Besides, in the framework of moduli stabilization in string compactifications --see e.g.\ the so-called large volume scenario~\cite{Balasubramanian:2005zx, Conlon:2005ki}--, the inflaton is often identified with a modulus field, meaning that it couples only gravitationally to the visible and hidden sectors. Therefore, reheating occurs at a slow pace via Planck-suppressed operators and the early phase of MD can have a prolonged duration. 

Gravitational collapse of overdensities and PBH formation during a MD epoch has already been studied;  see~\cite{Khlopov:1980mg} for pioneering work and \cite{Harada:2016mhb, Harada:2017fjm, Kokubu:2018fxy} for recent updates. Crucially, during MD any overdensity, no matter how small, undergoes gravitational collapse, in stark contrast to the case of collapse during RD. 
However, asphericities of the collapsing region strongly affect the outcome of the collapse: only very spherical overdensities may lead to PBHs \cite{Khlopov:1980mg, Harada:2016mhb}. Nonetheless, for not too small overdensities, the fraction of the energy density in the form of PBHs depends on the variance of the density fluctuations through a power-law, unlike in the RD case, where the dependence is exponential. In the case of MD, for small enough density fluctuations, the angular momentum of the collapsing region cannot be neglected  and the PBH fraction regains a multiplicative exponential factor~\cite{Harada:2017fjm}, but this exponential dependence is still much milder than in RD.   
Therefore, the enhancement of the primordial power spectrum required to form a significant amount of PBHs during MD can be orders of magnitude smaller than during RD, due to the absence of pressure.

In this work we show in a model-independent way that abundant PBH formation in the window \eq{win} is most effective for intermediate reheating temperatures $\lesssim 10^6$~GeV.\footnote{Similar conclusions were reached in previous works. In this paper, we expand with respect to~\cite{Harada:2017fjm} and improve previous estimates presented in~\cite{Dalianis:2018frf}, which also considers a modulated potential in a different inflationary setup.} For these temperatures, a power spectrum with an amplitude at small scales of order $10^{-4}$ {(or $10^{-3}$ depending on the width of the peak)} is enough to lead to a fraction of DM in the form of PBHs of order $1$, in contrast to the amplitude of order $10^{-2}$ required if the PBHs form during RD. This fact leads to a considerable reduction in the necessary tuning of the parameters of the inflaton potential, as we illustrate with concrete examples. Heavier PBHs can in principle also be formed in our scenario, but they require further tuning of the parameters in the inflationary potential, and lower reheating temperatures. Similarly, our setup also allows for PBH formation during RD, {with a higher cost} in terms of parameter tuning. 

{In the MD case, the milder fall of the collapsing energy density fraction away from the peak implies that a larger fraction of lighter PBHs is also formed. Therefore, the scenario which we consider in this work can be significantly constrained by PBH evaporation bounds. We discuss some possibilities for a viable scenario evading these bounds.}

This paper is structured as follows: in Section \ref{abundance}, we discuss the advantages of PBH formation during MD with respect to RD. We review the PBH abundance and mass formulae when they form during MD, and compare them to the case in which they form during RD. We also discuss the conditions under which a long epoch of MD due to inflaton oscillations can be achieved. In Section \ref{sec:potential}, we present the inflationary potential we consider, motivated by AMI. Section  \ref{results} is devoted to the numerical computation of the inflationary power spectrum for two examples. We discuss our findings in Section \ref{sec:conclusions}.

\section{Primordial black hole formation during matter domination}
\label{abundance}

PBHs can originate from the gravitational collapse of regions with large density fluctuations, which we assume are seeded by inflation. The mass and abundance of these PBHs depend on the equation of state of the Universe when the wave number of the fluctuations becomes comparable to the Hubble radius after inflation. During radiation domination (RD) the radiation pressure opposes the gravitational collapse, whereas during matter domination (MD) any overdensity grows since the pressure is zero. In this section we thus focus on PBH formation during an early epoch of MD, which starts right after the end of inflation. We use $t_{m}=2/(3H_{m})$ to denote the time at which MD ends. For simplicity, we consider that the Universe thermalizes instantaneously at $t_{m}$ and becomes radiation dominated. Given that in RD the Hubble expansion rate is $H=1/(2t)$  and the energy density during this period is therefore $\rho = 3 M_P^2/(4 t^2)$, we can define the temperature of the radiation bath at thermalization, $T$, through $\rho= (\pi^2/30)g(T)T^4$ as
\begin{equation}
\label{hubbleT}
T_{m}=\left(\frac{M_{p}}{t_{m}}\right)^{1/2}\left(\frac{4\pi^{2} g(T_m)}{90}\right)^{-1/4},
\end{equation}
where $g=\sum_b g_b +(7/8)\sum_f g_f$ counts the effective number of the degrees of freedom and the sums over $b$ and $f$ run, respectively, over the baryonic and fermionic species whose masses are below $T$ at any given time. In the Standard Model (SM), for temperatures above the electroweak phase transition one has $g=106.75$. 

\begin{figure}[t]
\includegraphics[scale=0.29]{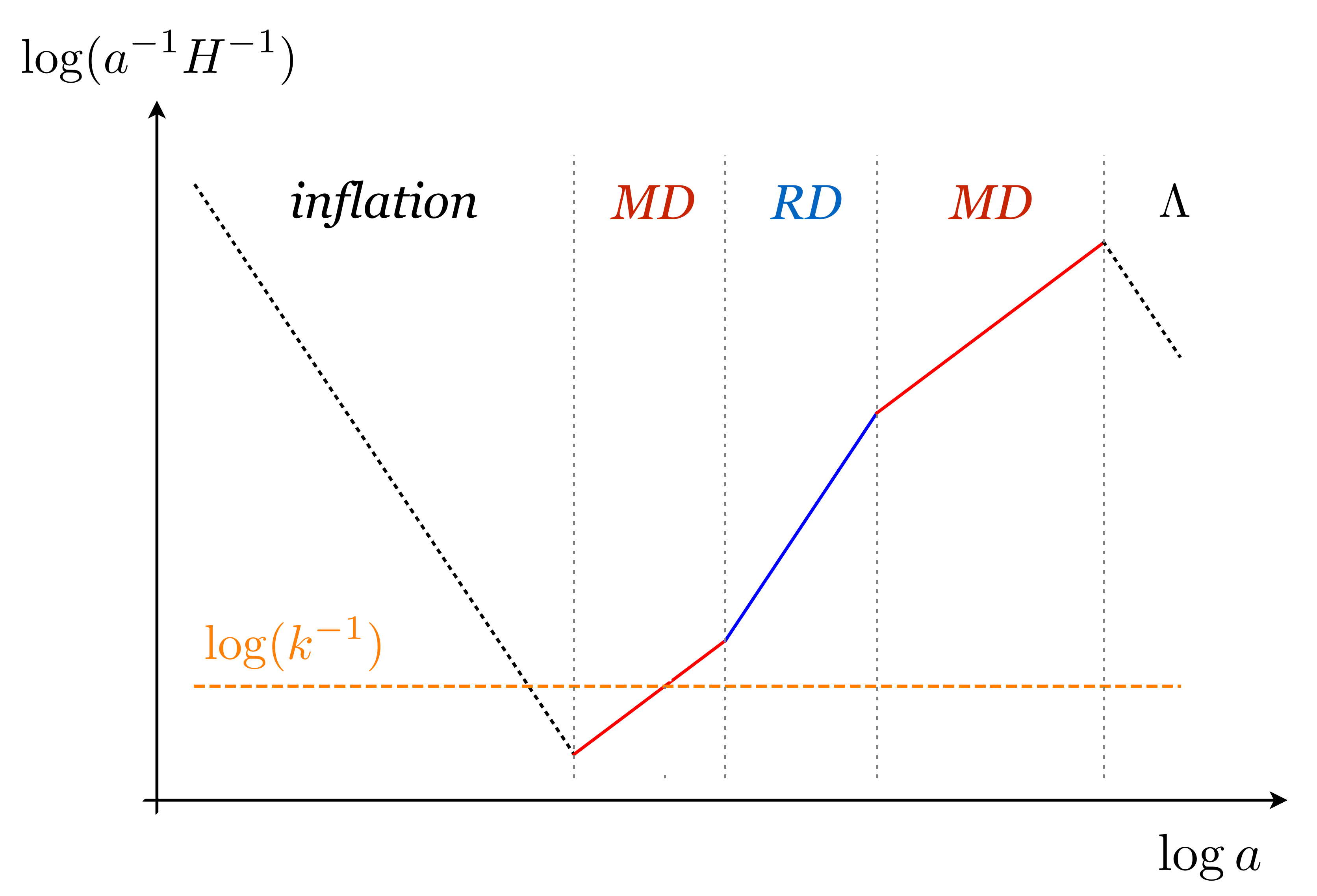} 
\centering
\caption{\it Schematic evolution of the inverse of the conformal Hubble function $\mathcal{H}=aH$ as a function of $a$, the scale factor of the Universe. The horizontal orange dashed line depicts a comoving distance scale that becomes of the size of a conformal Hubble patch (re-entering the horizon after inflation) during an early phase of matter domination.}
\label{fig:epochs}
\end{figure}

A schematic representation of the evolution of the scale factor across the various phases of the cosmological history in the scenario we consider is shown in Figure \ref{fig:epochs}. An early phase of MD after inflation can occur during perturbative reheating if the inflaton oscillates rapidly in a quadratic minimum \cite{Turner:1983he}.  In this case, we can identify the temperature $T_m$ with the reheating temperature and we will refer to it in this way through this work. We will elaborate more on the connection between $T_{m}$ and inflation in Section \ref{sub:reheating}. However, it is worth  noting that there are other possibilities to realize a phase of early MD (e.g.~oscillations of other heavy scalars) and our discussion until Section \ref{sub:reheating} does not depend on the origin of this phase. 

In the following subsections we review the expressions for PBH masses and abundances –assuming formation in either MD or RD– as functions of the primordial power spectrum, the comoving scale and the reheating temperature in the case of MD. We find that reheating temperatures $\lesssim 10^{6}~\text{GeV}$ are particularly interesting. In Section \ref{sub:reheating} we discuss in detail under which conditions such reheating temperatures can be obtained from inflaton oscillations and decay after inflation.

\subsection{Primordial black hole masses}
\label{sub:masses}

As mentioned above, PBHs form from the collapse of regions with large density fluctuations when their spatial extension, characterized by some scale $k$, becomes comparable to the size of a Hubble patch. The mass of the individual PBHs is mainly given by the Hubble mass at the time of horizon re-entry for the scale $k$. This scale cannot be  defined unambiguously; see \cite{Kalaja:2019uju} for a recent discussion. A common approximation for peaked spectra identifies $k$ with the location of the peak of the primordial  spectrum in linear perturbation theory. We will use this approximation, which is sufficient for our purposes. Then, the mass $M_{\mathrm{PBH}}$ of individual PBHs is
\begin{align}
\label{bhmass1}
M_{\mathrm{PBH}} = 4\pi\,\gamma\frac{M_p^2}{H}\,,
\end{align}
where  $H$ is equal to $1/(2t)$ for RD after inflation and $2/(3t)$ for a phase of early MD. The coefficient $\gamma$ quantifies the efficiency of the collapse.  Numerical analyses in the case of RD indicate that $\gamma$ depends on the spectral shape of the density fluctuations and that the actual mass depends mildly on the density threshold that triggers the formation of a PBH \cite{Niemeyer:1999ak}. We will neglect these dependencies and use $\gamma = 0.2$ for RD, see \cite{Carr:1975qj,Niemeyer:1999ak}. The actual efficiency of the collapse in MD is uncertain but may be expected to be higher than in RD due to the absence of radiation pressure. For concreteness, we take $\gamma =1$ for MD in the numerical examples of Section \ref{results}, although we keep $\gamma$ unspecified in most of the expressions below.

An overdensity of comoving scale $k$ re-enters the Hubble horizon at time $t_{k}$, when the condition $k=a(t_{k})H(t_{k})\equiv a_k H_k$ is satisfied. If this occurs during a MD phase which ends at time $t_{m}$, we can write: $a_{k}=\left({a_{k}}/{a_{m}}\right)\left({a_{m}}/{a_{0}}\right)$\,, where $a_{0}$ is the scale factor today, which we normalize to one.  Entropy conservation from $t_m$ until today implies that ${a_{m}}/{a_{0}}=(T_{0}/T_m)\left({g_s(T_0)}/{g_s(T_m)}\right)^{1/3}$, where $T_0$ is the current CMB temperature. During the phase of early MD there is no thermal equilibrium, but the scaling $a_k/a_m=(H_m/H_k)^{2/3}$ can be used. Combining these results with the condition for horizon crossing and using that $3M_p^2 H_m^2 =(\pi^2/30)g(T_m)T_m^4$ to eliminate $H_m$, we obtain
\begin{align}
a_k=\frac{\pi^2}{90}g(T_m)\frac{g_s(T_0)}{g_s(T_m)}\frac{T_0^3\,T_m}{k^2M_p^2}\,,
\end{align}
where we are keeping the number of effective entropy ($g_s$) and temperature ($g$) relativistic degrees of freedom distinct and $T_{0}$ is the temperature of radiation today. This expression allows us to write the PBH mass of \eq{bhmass1} as
\begin{align} \label{massmd2}
M_{\mathrm{PBH}}=\gamma\,\frac{2\pi^3}{45}\left(\frac{T_0}{k}\right)^3\frac{g_s(T_0)}{g_s(T_m)}\,g(T_m)\,T_m\,,\quad \text{for early MD}\,.
\end{align}
If the PBHs form during RD, the expression for their mass can be obtained following a similar logic. In this case
\begin{align}
a_k=\frac{\pi}{3\sqrt{10}}\left(\frac{g_s(T_0)}{g_s(T_k)}\right)^{2/3}\frac{T_0^2}{k\,M_p}\sqrt{g(T_k)},
\end{align}
and therefore
\begin{align}
M_{\mathrm{PBH}}=\gamma\,\frac{4\pi^2}{3\sqrt{10}}\left(\frac{T_0}{k}\right)^2\left(\frac{g_s(T_0)}{g_s(T_k)}\right)^{2/3}\sqrt{g(T_k)}\,M_p\,\quad \text{for RD}\,.
\end{align}
The PBH mass thus scales as $k^{-2}$ if  PBHs form during RD and $k^{-3}$ during early MD. In the  latter case the PBH mass depends also on the duration of the phase of early MD through the reheating temperature $T_m$, see \eqref{hubbleT}. For the purpose of comparison, it is useful to write both mass expressions in terms of some benchmark values for $k$, $T_m$, and the mass of the Sun, $M_\odot$:
\begin{align}
\label{massrd}
M_{\mathrm{PBH}}&\simeq 2.8\cdot 10^{-16}~\left(\frac{\gamma}{0.2}\right)\left(\frac{g_{}(T_{k})}{g_{ s}(T_{k})}\right)^{2/3}\left(\frac{106.75}{g_{}(T_{k})}\right)^{1/6}\left(\frac{10^{14}~\text{Mpc}^{-1}}{k}\right)^{2}M_{\odot}\quad \text{for RD},\\
\label{massmd}
M_{\mathrm{PBH}}&\simeq 2.4\cdot 10^{-17}~\gamma\left(\frac{g_{}(T_{m})}{g_{ s}(T_{m})}\right)\left(\frac{10^{14}~\text{Mpc}^{-1}}{k}\right)^{3}\left(\frac{T_{m}}{10^{5}~\text{GeV}}\right)M_{\odot}\quad \text{for early MD}\,.
\end{align}
{The expressions above have been obtained by setting $T_0=2.7255$~K~\cite{2009ApJ...707..916F}, $g(T_0)=2.00$, and $g_{s}(T_{0})=3.91$. These values for the entropy and temperature degrees of freedom correspond to assuming that all three neutrinos are non-relativistic today, see~\cite{Kalaja:2019uju}.} We plot equation~\eqref{massmd} in Figure \ref{abundancemd} (solid orange lines), together with other quantities and constraints which we introduce in the following subsections. 

\subsection{Primordial black hole abundance}

We are interested in the current abundance of PBHs with respect to that of DM:
\begin{align}
f_{\rm PBH}=\frac{\Omega^0_{\text{PBH}}}{\Omega^0_{\text{DM}}}\,.
\end{align}
In the approximation of rapid collapse of the overdensity, $f_{\mathrm{PBH}}$ can be written in terms of $\beta$, i.e.~the ratio of the collapsing energy density to the total energy density at the time of the collapse:
\begin{align}
\beta =\frac{1}{\gamma}\frac{\rho_{\text{PBH}}(t_k)}{\rho(t_k)}\,.
\end{align}
Here $\rho_{\text{PBH}}$ and $\rho$ are the PBH and total energy densities, respectively. As in the previous subsection, the constant $\gamma$ encodes the efficiency of the collapse, see \eqref{bhmass1}. 

In the case of PBHs formed during an early phase of MD, by means of entropy conservation and using $\rho_{m}=(\pi^2 g(T_{m})/30) T^{4}_{m}$, we obtain
\begin{align}
\label{eq:frac1} f_{\mathrm{PBH}}=\gamma\,\beta\,\frac{\Omega^0_\gamma}{\Omega^0_{\rm DM}}\,\frac{g(T_m)g_s(T_0)}{g(T_0)g_s(T_m)}\,\frac{T_m}{T_0}\,.
\end{align}
As explained before, to obtain this result we have assumed that the transition between the different epochs depicted in Figure \ref{fig:epochs} is instantaneous.

The analogous expression for RD is obtained from \eqref{eq:frac1} by simply setting $T_{m}=T_{k}$, where $T_k$ is the temperature of the radiation at the time of formation. In this case we can write $T_k$ as a function of the Hubble rate and relate this to the PBH mass through \eq{bhmass1}. Then, the expressions for the PBH abundance as a function of the quantity $\beta$  in the RD and early MD cases are, respectively:
\begin{align}
\label{fracrad}
f_{\mathrm{PBH}}&\simeq \left(\frac{\gamma}{0.2}\right)^{3/2}\left(\frac{\beta}{8.9\cdot 10^{-16}}\right)\left(\frac{g_{}(T_{k})}{106.75}\right)^{-1/4}\left(\frac{g(T_{k})}{g_{s}(T_{k})}\right)\left(\frac{M_{\mathrm{PBH}}}{10^{-15}~M_{\odot}}\right)^{-1/2}\quad \text{for RD}\,,\\
\label{fracmat}
f_{\mathrm{PBH}}&\simeq \gamma \left(\frac{\beta}{5.5\cdot 10^{-15}}\right)\left(\frac{g(T_{m})}{g_s(T_m)}\right)\left(\frac{T_{m}}{10^{5}~\text{GeV}}\right)\quad \text{for early MD}\,.
\end{align}
The temperature dependence of the PBH abundance in the early MD case implies that a shorter duration of this phase (i.e.\ a higher reheating  temperature) implies a larger abundance, see equation \eq{hubbleT}. This is simply due to the fact that PBHs, being cold dark matter, dilute slower than radiation as the Universe expands. Therefore, the longer the duration of the RD phase is (i.e.\ the shorter is the early MD phase), the higher the abundance of PBHs. 

Notice that we could also write $f_{\mathrm{PBH}}$ in the early MD case as a function of the PBH mass, using \eq{massmd}. However, unlike in the RD case of \eq{fracrad} this introduces explicitly the wavenumber $k$, which makes the formula more cumbersome. 

{In the previous expressions we have assumed that the distribution of PBH masses at formation time is monochromatic. This is a good approximation for sufficiently peaked primordial spectra. If this condition is not satisfied,  as will be the case in our numerical examples, the total PBH abundance can be obtained by integrating $\int f_{\mathrm{PBH}}(M_\mathrm{PBH})d\log(M_\mathrm{PBH})$.}

\subsection{The collapsing fraction of the energy density}

So far we have obtained general expressions for the abundance as functions of the collapsing energy density fraction $\beta$, but we have not discussed how this (model dependent) quantity is computed in each case. In the RD case we use the approximation
\begin{align}
\label{betard}
 \beta(k)&= \frac{1}{\sqrt{2\pi\sigma^2(k)}}\int_{\delta_c}^\infty\exp\left[-\frac{\delta^2}{2\sigma^2(k)}\right]d\delta
 ,\quad \text{for RD}.
\end{align}
In this expression $\delta=\delta\rho/\rho$ is the density contrast in the total matter gauge (see \cite{Ballesteros:2018wlw}), $\delta_c$ is the estimate for threshold on $\delta$ for gravitational collapse during RD and $\sigma^2(k)$ is the variance of the density contrast smoothed over a comoving distance scale $\sim 1/k$, given by 
\begin{equation}\label{beta1}
\sigma^2(k)=\frac{4(1+w)^2}{(5+3w)^2}\int\frac{dq}{q}\Big(\,\frac{q}{k}\,\Big)^4 \mathcal{P}_{\mathcal{R}}(q) W^2(q/k)\,,
\end{equation}
where $w=1/3$ for RD. In this expression $\mathcal{P}_{\mathcal{R}}(k)$ is the dimensionless power spectrum of the comoving curvature perturbation \(\mathcal{R}\)  and \(W(x)\) is a window function which we will take to be Gaussian. The expression \eq{betard} is obtained by applying the Press-Schechter formalism \cite{Press:1973iz} and assuming that the primordial fluctuations leading to PBH formation are Gaussian. The value of $\delta_c$ is known to depend on the profile of the collapsing overdensities, but should be between $0.4$ and $0.5$, see \cite{Musco:2004ak,Musco:2008hv,Musco:2012au }. 

The physical interpretation of \eq{betard} is transparent: only overdensities above the threshold $\delta_c$ can collapse into a black hole. For PBHs formed during MD, the situation is very different, due to the absence of radiation pressure. In  this case, the equation \eqref{beta1} still applies (now with $w=0$) but, for sufficiently large variances, the collapsing energy fraction has been estimated to be \cite{Khlopov:1980mg,Harada:2016mhb}:\footnote{Inhomogeneities of the collapsing overdensity may further suppress PBH formation during MD~\cite{Kokubu:2018fxy}. However, such extra suppression depends on certain assumptions on the final stages of the collapse, which might be evaded in realistic setups; see the discussion in~\cite{Kokubu:2018fxy}. For these reasons, we neglect inhomogeneities in our estimates and use \eqref{betamd} and \eqref{betaam} throughout this work.}
\begin{align}
\label{betamd}
\beta(k)& \simeq 0.056\,\sigma^{5}(k),\quad \text{for early MD, and}\,\,\sigma\gtrsim\sigma_{\mathrm{ang}}.
\end{align}
This expression for $\beta$, which accounts for the effect of asphericities in the collapsing region,  is valid only if $\sigma$ is larger than a certain value $\sigma_\text{ang}\simeq0.005$  \cite{Harada:2017fjm}. Below this value, the effect of the angular momentum of the collapsing region becomes relevant, and the expression above must be replaced by~\cite{Harada:2017fjm}:
\begin{align}
\label{betaam}
\beta(k)& \simeq 1.9\times10^{-7}f_q(q_c)\mathcal{I}^6\sigma^2(k)\exp\left[-0.147\left(\frac{\mathcal{I}^{2}}{\sigma(k)}\right)^{2/3}\right],\quad \text{for early MD, and}\;\; \sigma\lesssim \sigma_{\text{ang}}.
\end{align}
Here, $\mathcal{I}$ is an $\mathcal{O}(1)$ parameter\footnote{The variance of the angular momentum $\langle \mathbf{L}^2\rangle$ and $\sigma$ are related through $\mathcal{I}$ as follows: $45\, t\, \langle \mathbf{L}^2\rangle^{1/2} \simeq 8\pi (a\,r)^5 \rho\, \mathcal{I}\, \sigma$, where $r$ is the initial comoving radius of the overdensity and $\rho$ is the homogeneous energy density  of the MD universe.}\cite{Harada:2017fjm} and $f_q(q_c)$ is the fraction of mass with a level of quadrupolar asphericity $q$ smaller than a threshold $q_c\simeq 2.4 (\mathcal{I}\, \sigma)^{1/3}$. Following the estimates of \cite{Harada:2017fjm} we will assume $f_q(q_c)=1$ in our numerical examples.

\begin{figure}[t]
\centering
\includegraphics[width=.45\textwidth]{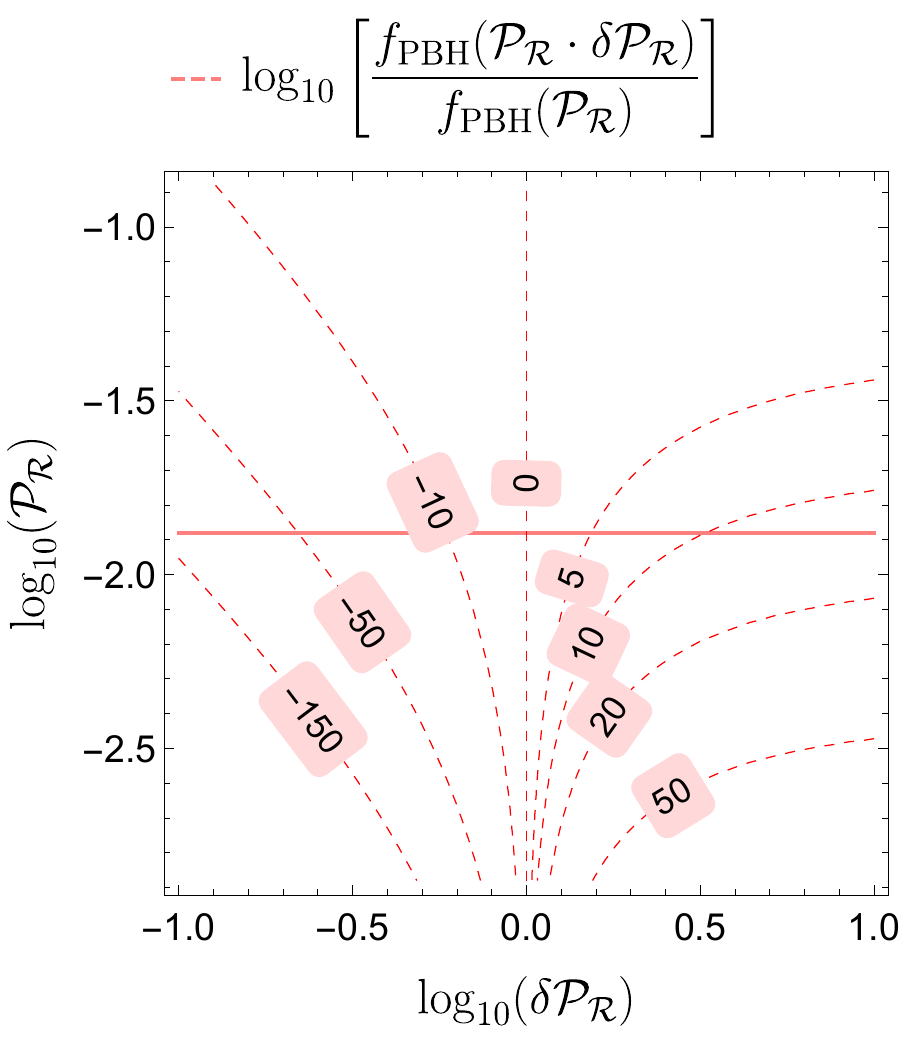}\hfill
\includegraphics[width=.45\textwidth]{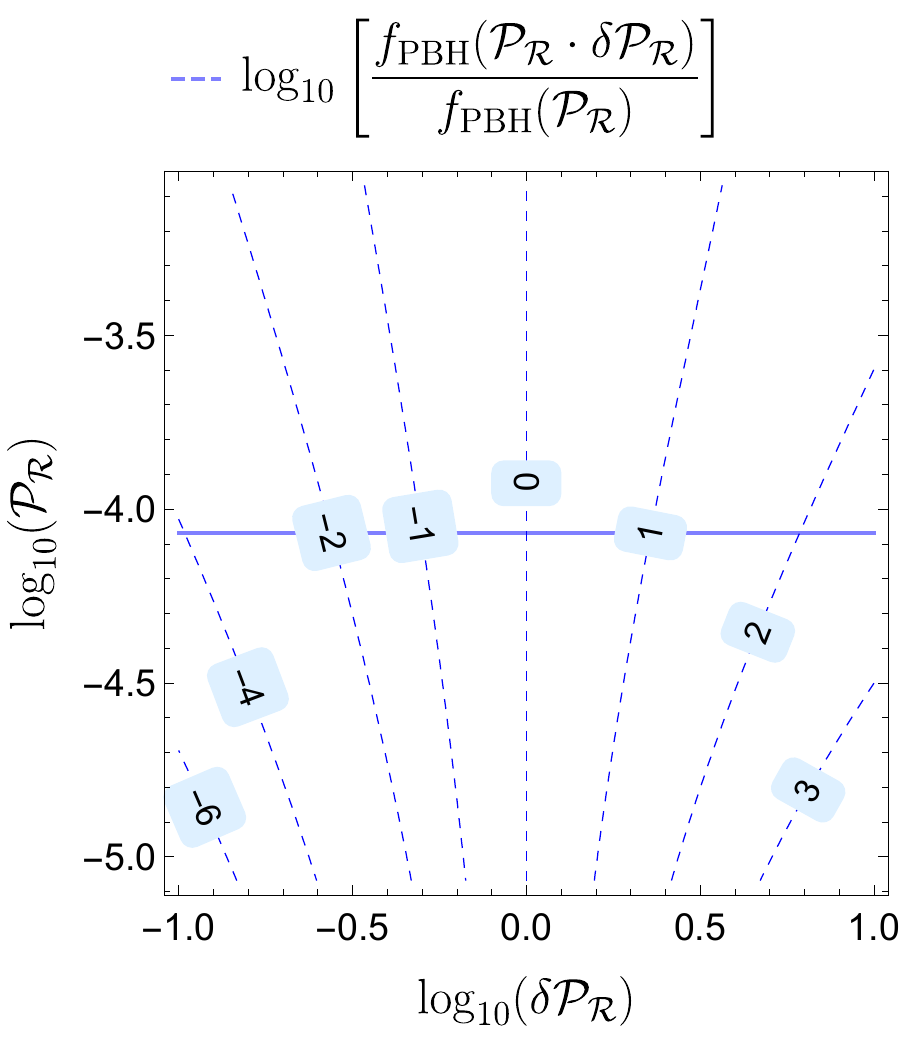}
\caption{\it Sensitivity of the fraction of DM in the form of PBHs to variations in the power spectrum for RD (equation \eq{betard}, left panel) and MD with angular momentum effects (equation \eq{betaam}, right panel). The dashed lines represent the relative change in $f_{\mathrm{PBH}}$ when the power spectrum is varied with a multiplicative factor $\delta \mathcal{P_R}$, and the horizontal (solid) lines represent the value of the primordial power spectrum required to get $f_{\mathrm{PBH}}=1$ in each case. We have used \(M_{\mathrm{PBH}}=10^{-15}M_\odot\), \(\gamma=0.2\), and approximated \(\sigma^2\simeq(16/81)\mathcal{P}_\mathcal{R}\) for RD (left panel), and \(T_m=10^5\mathrm{GeV}\), \(\gamma=1\), $f_q(q_c)=1$, $\mathcal{I}=1$, and \(\sigma^2\simeq(4/25)\mathcal{P}_\mathcal{R}\) for MD (right panel). An order of magnitude change in $\mathcal{P_R}$ has a much greater impact on $f_{\mathrm{PBH}}$ in RD than in MD.}
\label{fig:tuning}
\end{figure}

The previous equations summarize a key difference between PBH formation in MD and RD. Whereas in RD the PBH abundance is exponentially sensitive to the primordial power spectrum $\mathcal{P_R}$, in the MD case this dependence is a power-law for fluctuations larger than $\sigma_{\text{ang}}$. The reason for this difference lies in the threshold for gravitational collapse: in RD this is given by $\delta_{c}$, while in MD essentially any overdensity undergoes gravitational collapse, due to the very small Jeans length  of non-relativistic matter.\footnote{The Jeans length determines the critical radius above which an overdensity collapses and is proportional to the  speed of sound of the fluctuations.} Angular momentum effects need to be taken into account in MD if $\sigma$ is small enough. In that case, the approximate power-law behavior of \(\beta\) is lost, but its sensitivity to the primordial power spectrum is still much milder than in RD. In Figure \ref{fig:tuning} we compare the sensitivity of $\beta$ to changes in $\mathcal{P_R}$. The PBH fraction changes much more dramatically  with $\mathcal{P_R}$ for PBHs formed during RD than in MD. As we will show quantitatively in Section \ref{results}, this translates into a higher level of tuning in the  parameters of the inflationary potential if PBHs with  $f_{\rm PBH}\sim 1$ form during  RD.

\begin{figure}[t]
\centering
\includegraphics[width=.4\textwidth]{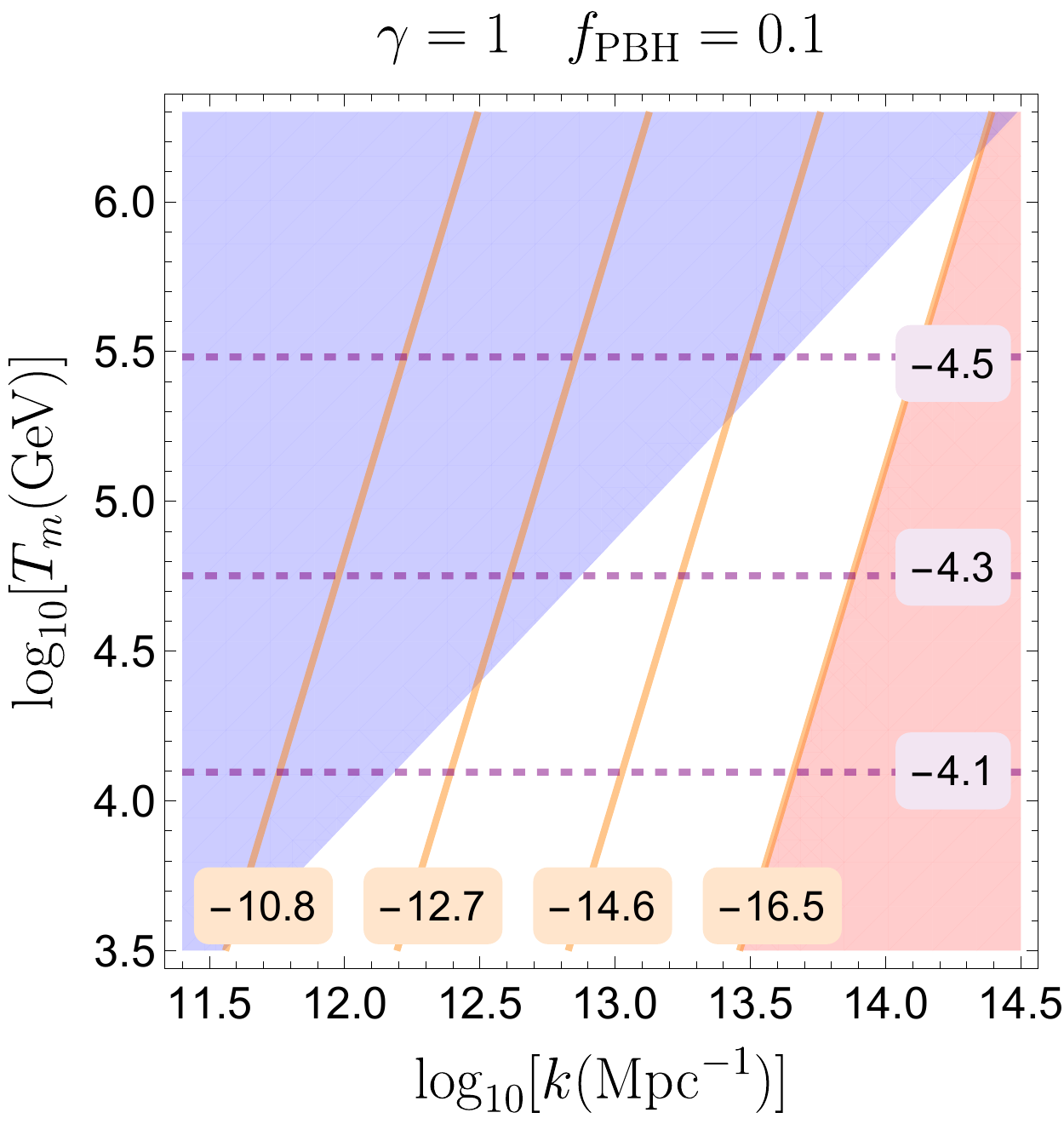}\hfill
\includegraphics[width=.6\textwidth]{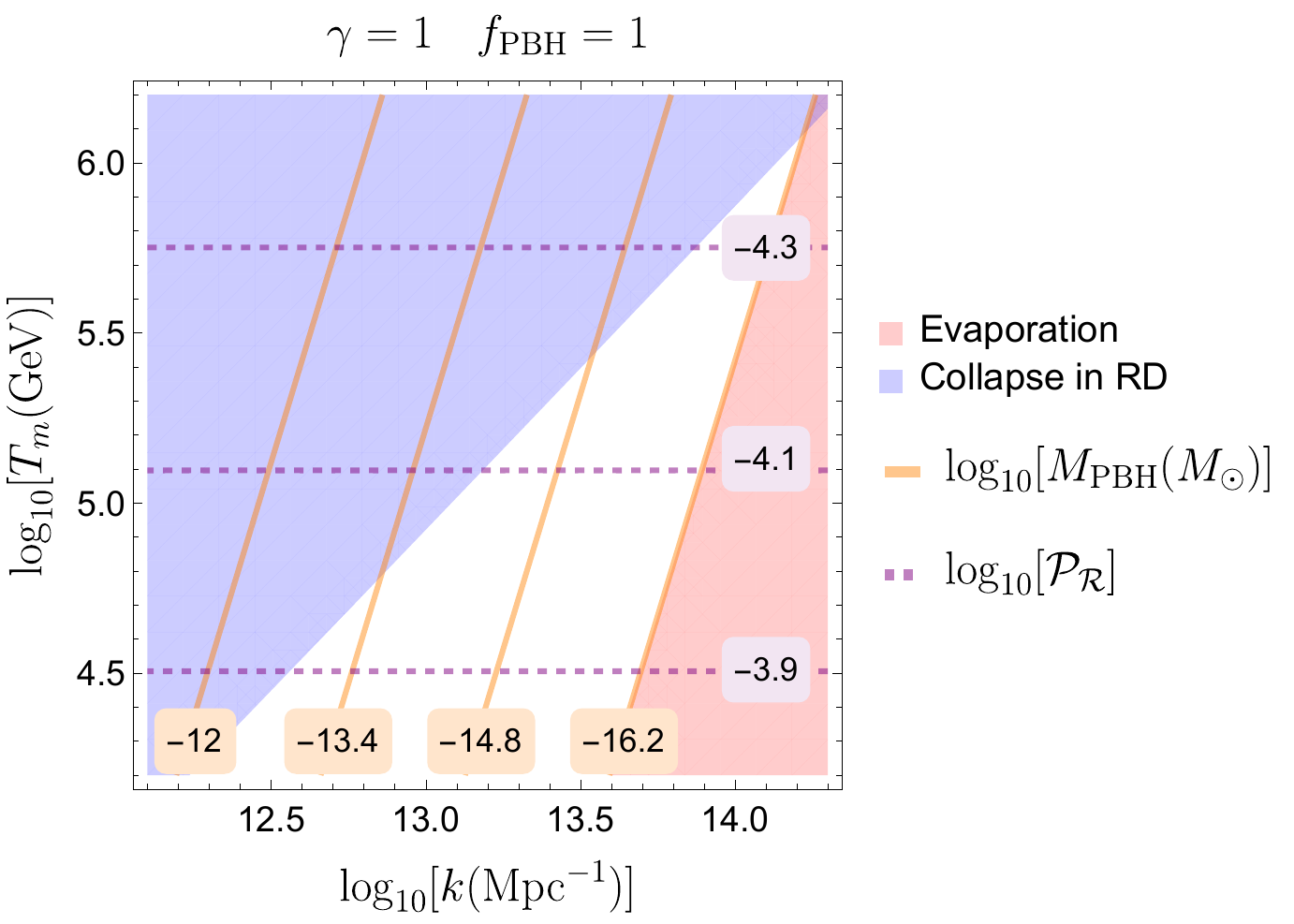}
\caption{\it Mass of PBHs formed during an early matter dominated phase as a function of the scale \(k\) of the collapsing fluctuation and the reheating temperature \(T_m\), fixing \(\gamma=1\), for \(f_{\mathrm{PBH}}=0.1\) (left) and \(f_{\mathrm{PBH}}=1\) (right). The dashed purple lines show the height of the power spectrum required to get the fixed value of \(f_{\mathrm{PBH}}\) as a function of \(T_m\). We use the formula for the fraction~\eqref{betaam}, which is the relevant one in the plotted region of parameter space, with $f_q(q_c)=\mathcal{I}=1$, and the approximation \(\sigma^2\simeq(4/25)\mathcal{P}_\mathcal{R}.\) PBHs with masses below \(\sim10^{-16}M_\odot\) are evaporating today and constrained by extragalactic gamma-rays \cite{Carr:2009jm, Ballesteros:2019exr, Arbey:2019vqx} (pink-shaded region). Similarly, masses above \(\sim 10^{-11}M_\odot\) are constrained by microlensing see \cite{Niikura:2017zjd, Montero-Camacho:2019jte} (not shown here). The blue-shaded region corresponds to the constraint \eqref{constraintwithl}.}
\label{abundancemd}
\end{figure}

Equations \eqref{betard} and \eqref{betamd} should be supplemented with an additional constraint which arises from requiring that the collapsing fluctuation reaches the non-linear regime during the MD era. This effect was neglected in~\cite{Dalianis:2018frf}, which assumed instantaneous collapse. However, the above requirement significantly limits the available parameter space, which  we show in Figure \ref{abundancemd}. Using that the linear density contrast grows as the scale factor during matter domination, one finds that only fluctuations larger than $\sigma_{\text{nl}}\simeq (H_{m}/H_{k})^{2/3}$ reach non-linearity before the end of the early MD epoch. Smaller fluctuations take longer to reach the non-linear regime and thus do not complete the collapse during MD. This constraint can be conveniently expressed as follows:
\begin{equation}
\label{constraintnol}
\sigma\gtrsim1.9\cdot10^{-4}\left(\frac{g(T_m)}{106.75}\right)\left(\frac{106.75}{g_s(T_m)}\right)^{2/3}\left(\frac{T_m}{10^5\mathrm{GeV}}\right)^2\left(\frac{10^{14}\mathrm{Mpc}^{-1}}{k}\right)^2.
\end{equation}
When the rotation of the collapsing fluctuation plays a role (i.e.\ for $\sigma\lesssim \sigma_{\rm ang}$) the constraint is slightly stronger than \eqref{constraintnol}. This can be understood as follows: during the linear evolution of an overdensity, its angular momentum grows; in particular, the longer the duration of the linear evolution, the stronger will be the effect of angular momentum on the gravitational collapse. Therefore, a different lower bound on $\sigma$ arises from requiring that the growth of angular momentum does not prevent PBH formation. The resulting constraint is $\sigma\gtrsim 5H_{m}/(2\,\mathcal{I}\,H_{k})$ --we refer the reader to~\cite{Harada:2017fjm} for its derivation--  and can be rewritten as
\begin{equation}
\label{constraintwithl}
\sigma\gtrsim10^{-5}\left(\frac{1}{\mathcal{I}}\right)\left(\frac{g(T_m)}{106.75}\right)^{3/2}\left(\frac{106.75}{g_s(T_m)}\right)\left(\frac{T_m}{10^5\mathrm{GeV}}\right)^3\left(\frac{10^{14}\mathrm{Mpc}^{-1}}{k}\right)^3.
\end{equation}
The advantage of considering an early phase of MD for PBH formation is reflected in Figure \ref{abundancemd}, where we plot the PBH masses according to \eqref{massmd} (solid orange lines) as well as the amplitude of the primordial power spectrum required to obtain $f_{\mathrm{PBH}}=1$ (dashed purple lines) combining \eqref{fracmat} and \eqref{betaam}, which is the appropriate expression for $\beta$ in the region of parameter space where $\sigma\lesssim \sigma_{\mathrm{ang}}$.\footnote{The analysis of \cite{Harada:2017fjm} suggests that the value $\sigma_{\mathrm{ang}}\simeq0.005$ should be taken as an order of magnitude estimate, rather than as a sharp threshold. In particular, effects of order higher than second in angular momentum may lower this value of $\sigma_{\mathrm{ang}}$ and extend the validity of \eqref{betamd}. This would lead to slightly smaller values of $\mathcal{P_R}$ being required for $f_{\rm  PBH}$ smaller than one, which are more advantageous for PBH formation, but the constraint \eqref{constraintnol} would then disfavor a larger region of parameter space, leading to slightly lower reheating temperatures.}  {We also show the observational Hawking evaporation bounds derived from the extra-galactic gamma-ray background assuming zero spin and a monochromatic mass distribution} \cite{Carr:2009jm,Arbey:2019vqx,Ballesteros:2019exr} (shaded pink region; {see Section~\ref{results} for a discussion on this bound}) as well as the constraint~\eqref{constraintwithl} (shaded blue region). In the viable region of parameter space, the minimal amplitude of the primordial power spectrum which is needed to obtain $f_{\mathrm{PBH}}=1$ is $\mathcal{P_R}\simeq10^{-4}$. This is five orders of magnitude above the value at CMB scales and two orders of magnitude smaller than the benchmark value required in the case of RD. Figure \ref{abundancemd} shows that $\mathcal{P_R}\simeq10^{-4}$ corresponds to a value for the reheating temperature $T_{m}\lesssim 10^{6}~\text{GeV}$, and $k\simeq 10^{14}$~Mpc$^{-1}$, which implies $M_{\text{PBH}}\sim 10^{-16}~M_{\odot}$, close to evaporation constraints. Values of $T_m$ and $k$ an order of magnitude away from these ones lead to PBH masses well within the currently allowed region.

\subsection{A long matter era from perturbative reheating}
\label{sub:reheating}

In the previous subsection we found that intermediate reheating temperatures $\lesssim 10^6~\text{GeV}$ are required to take advantage of an early phase of MD and produce PBHs of masses in the unconstrained window \eq{win} with $f_{\rm PBH}\sim 1$. While we did not specify the origin of the MD phase, a straightforward option exists in the context of inflation. The energy density of the inflaton scales as that of matter after inflation, if the inflaton undergoes small oscillations in	an approximately  quadratic minimum~\cite{Turner:1983he}. We will now discuss under which conditions a long epoch of MD due to oscillations of the inflaton can be realized, with resulting values of $T_{m}\lesssim 10^6$ GeV.

A perturbative description of reheating is not appropriate in general, due to the occurrence of preheating (see e.g.\cite{Kofman:1997yn}). This is the process by which explosive particle production takes place as a consequence of parametric resonance driven by the inflaton oscillations. The occurrence of preheating typically quenches the existence of a prolonged phase of early MD~\cite{Podolsky:2005bw}. Therefore, it is worth recalling under which conditions preheating is prevented. We illustrate this by considering a coupling of the form $\mu\, \phi\, \chi^2$ between the inflaton $\phi$ and another scalar field $\chi$.  If the inflaton potential is approximated around its minimum as $m^2\phi^2/2$, then $\phi$ undergoes damped oscillations after inflation due to  Hubble  friction. Approximately, $\phi(t)\simeq \Phi(t) \sin(m\,t) = M_{p}/(\sqrt{3\pi}mt)\sin(m\,t)$, being $m\,t\simeq n/2\pi$, where $n$ is the number of accumulated oscillations until the time $t$. By studying the growth of $\chi$ fluctuations, efficient preheating can be shown to occur as long as both of the following conditions are fulfilled~\cite{Kofman:1997yn}:
\begin{align} \label{precond}
\mu \lesssim 32\, \Phi(t),\quad {\rm and}\quad  4\mu\, \Phi(t) \gtrsim m^{3/2} H^{1/2}\,.
\end{align}
Violating one of these two conditions is enough to ensure that preheating does not happen. The first of these inequalities will be satisfied initially (for small $n$) provided that $\mu \ll M_p$. However, the second condition can be violated if $\mu$ is sufficiently small, preventing preheating from occurring initially. Once $n$ is large enough, both conditions are violated and preheating never occurs. 

Let us therefore impose that preheating does not occur and derive the value of $\mu$ such that $T_{m}\lesssim 10^{6}~\text{GeV}$ can be achieved. For the reheating channel under consideration, the perturbative decay rate of $\phi$ into $\chi$ is given by $\Gamma = \mu^2/(8\pi\, m)$. In the absence of preheating, perturbative reheating proceeds until $H\sim \Gamma$, when the energy density stored in the inflaton is approximately $3 M_p^2\Gamma^2$. Equating this to the energy density of the radiation bath (under our assumption of instantaneous transition between MD and RD), we get
\begin{align} \label{tempgamma}
T_{m}=\sqrt{\Gamma M_{p}}\left(\frac{\pi^{2}g(T_{m})}{90}\right)^{-1/4}\simeq\; 10^{5}~\text{GeV}\left(\frac{\mu}{10^{3}~\text{GeV}}\right)\left(\frac{10^{13}~\text{GeV}}{m}\right)^{1/2}.
\end{align}
Therefore, $T_{m}\sim\mathcal{O}(10^{5})~\text{GeV}$ is achieved if $\mu\sim 10^{-16}~M_{p}$ for typical values of the inflaton mass in high scale models, $m\sim10^{13}~\text{GeV}$. It is straightforward to check that for such small values of $\mu$ the second of the conditions \eqref{precond} is violated, which shows that the estimate \eqref{tempgamma} is consistent with the assumption of inefficient preheating.

It remains to be seen whether such small values of $\mu$ are feasible in concrete scenarios of inflation and reheating. Interestingly, in inflationary models inspired by string compactifications, such as the one that we will consider in the next section, the inflaton can be a modulus and thus couple only gravitationally to light degrees of freedom. In this case, the decay rate $\Gamma$ is Planck-suppressed, its typical form for decay into a scalar pair being $\Gamma\sim {m^{3}}/(48\pi M_{p}^{2})$~(see e.g.~\cite{Cicoli:2012aq}). For $m\sim 5 \cdot 10^{-6} M_p$ this translates into $\mu\sim 10^{-11}M_{p}$. For these values of $\mu$, preheating is again avoided. However, reaching $T_{m}\lesssim 10^{6}~\text{GeV}$ still requires an extra suppression of the decay rate. Model building possibilities in this direction can be found e.g.\ in~\cite{Cicoli:2012aq}, in the framework of the large volume scenario~\cite{Balasubramanian:2005zx, Conlon:2005ki} of moduli stabilization.

For the potential that we consider in Section \ref{sec:potential}, $m$ turns out to be about an order of magnitude larger than in the simple quadratic case. In this case, a stronger suppression of $\Gamma$ is needed for $T_m\lesssim 10^6$~GeV. This may be achieved in concrete realizations \cite{Cicoli:2012aq} of our inflationary setup. Alternatively, the early phase of MD can be extended by (or be entirely due to) extra heavy scalars with Planck-suppressed $\Gamma$ that start oscillating after the inflaton field decays. This kind of scenario can arise in string compactifications \cite{Banks:1993en,deCarlos:1993wie}. See also \cite{Dalianis:2018frf} for similar considerations applied to PBHs.

Before moving on to present the inflationary potential we will consider, let us relate the reheating temperature to the number of e-folds between the time at which the largest observable scales left the horizon and the end of inflation. The number of e-folds ($N=\int H dt$) elapsed from the moment a scale $k$ satisfies $k = a\,H$ during inflation until its end is:\footnote{A similar expression was first given in \cite{Liddle:2003as}. We have followed an analogous derivation and chosen to  write the numerical factors differently.}
\begin{align}
\label{totalefolds}
N(k)=63.55+\frac{1}{4}\log\frac{\Omega_r^{0}}{h^2}-\log\frac{k}{a_0H_0}-\frac{1}{12}\log\frac{\rho_{end}}{\rho_m}+\frac{1}{4}\log\frac{\rho_k}{\rho_{end}}+\frac{1}{4}\log\frac{\rho_k}{(10^{16} \rm{GeV})^4}\,,
\end{align}
where $H_0=100\,h\, \rm{km\, s^{-1} Mpc^{-1}}$ and $\rho_k$ is the energy density of the universe at $k=a\,H$ during inflation. The subscripts $_{end}$ and $_{m}$ refer to the end of inflation and the end of the period of early MD. The quantity $\Omega_{r}^{0}$ is the current radiation density. We can relate $\rho_{m}$ to $T_m$ simply through
\begin{align} \label{denst}
\rho_{m}=\frac{\pi^2}{30}g(T_m)T_m^4\,.
\end{align}
Given a model of inflation, we can determine $\rho_k$ and $\rho_{end}$  and then use the last two equations to find out the required reheating temperature $T_m$ for a specific value of $N(k)$. For instance, if we assume $10~H_{end}\sim H_k\simeq 10^{13}$GeV, so that $10^2\rho_{end}\sim\rho_k\sim (10^{16}\rm GeV)^4$, we get
\begin{equation} \label{hexit}
N(k)\simeq 49-\log\left(\frac{k}{0.05~\text{Mpc}^{-1}}\right)+\frac{1}{3}\log\left(\frac{T_{m}}{10^{5}~\text{GeV}}\right)\,,
\end{equation}
where we have used the values of the cosmological parameters listed in appendix \ref{cosmoparams}. If the reheating temperature is $T_{m}\sim 10^{5}~\text{GeV}$, inflation lasts approximately 50 e-folds after fluctuations of wavenumbers comparable to the Planck fiducial scale ($k=0.05$ Mpc$^{-1}$) exit the horizon. As we have seen in the previous subsection, the most interesting scales for PBH formation during MD are around $k\sim 10^{14}~\text{Mpc}^{-1}$. According to \eq{hexit}, these fluctuations should exit the horizon during inflation approximately 15 e-folds before the end of inflation. In  Section~\ref{results} we will see how these observations translate into constraints on the parameters of the large-field inflationary potential that we introduce in the next section. 

\section{Inflation from axion monodromy}
\label{sec:potential}

{{We emphasize that the  discussion of Section~\ref{sub:reheating} is equally valid for large- and small-field inflation models, since the only properties we are  invoking are a quadratic minimum --the potential may have a different form away from it-- and weak couplings of the inflaton to other species. Since small-field models allow to implement low-scale inflation more easily it may be natural to think they are better suited  to accommodate a long phase of matter domination after inflation (with a small reheating temperature), but this may not  be necessarily the case. Let us illustrate this with a well-known example: the hilltop-like potential $V=V_0(1-(\phi/\phi_0)^p)^2$, where $p$ is a rational number. This potential features an  approximately quadratic minimum for small deviations away from $\phi=\phi_0$. However, requiring $n_s\simeq 0.96$ and $H\ll 10^{10}$~GeV, we find that $\sim 5$ oscillations after the end of inflation have to occur before reaching a discrepancy below the $\sim 10\%$ level  between the potential and its quadratic approximation. In contrast, the large-field model we will now introduce differs from its quadratic approximation around the minimum by at most $\sim 1\%$ already at the end of inflation. It thus seems that  implementing in small-field models the sort of potential features required for abundant PBH production  is more cumbersome than for large-field models and probably requires a higher degree of tuning, and we will not consider this possibility any further.}}

In the  context of canonical single-field inflation, the basic feature that a potential should have in order to lead to PBH formation is the presence of either a near-inflection point or a local minimum followed by a local maximum. Such an ingredient is certainly non-generic for inflationary potentials; the great majority of existing models of PBHs make use of potentials with coefficients chosen in such a way as to generate the desired stationary points, without addressing the question of why such points should exist in the first place. 

Here we provide a well-motivated inflationary potential which naturally features stationary points as a consequence of an underlying symmetry (see~\cite{Ozsoy:2018flq} for a previous work in this direction). We focus on a scenario in which the inflaton is originally identified with a pseudo-Goldstone boson originating from a continuous shift symmetry, broken to a discrete subgroup by non-perturbative effects \cite{Freese:1990rb}. {Such a field is commonly referred to as an \emph{axion}, in analogy with the familiar case of QCD~\cite{Peccei:1977hh, Wilczek:1977pj, Weinberg:1977ma}. Axions are common inflaton candidates, especially in string-inspired scenarios. See e.g.~\cite{Baumann:2014nda} for an extensive review.}

In the latter context, a family of well-motivated axion potentials arises, where the potential energy of the inflaton varies as the field completes a cycle in its originally compact field space. These constructions are known as axion monodromy inflation (AMI)~\cite{Silverstein:2008sg, McAllister:2008hb}. The inflationary potentials belonging to this class typically exhibit the following functional form:
\begin{equation}
\label{eq:monpot}
V(\phi)=V_{\text{mon}}+V_{\text{cos}}=\frac{m^{2}F^{2}}{2p}\left[-1+\left(1+\frac{\phi^{2}}{F^{2}}\right)^{p}\right]+\Lambda(\phi)^{4}\cos\left(\frac{\phi}{f}+\delta\right).
\end{equation}

Let us first focus on $V_{\text{mon}}$. This term features three parameters: two energy scales, $F$ and $m$, and an exponent $p$, which may be either positive or negative. For $\phi\ll F$, this part of the potential is well approximated by a parabola. For $\phi\gg F$, the potential either grows as $\phi^{2p}$ (if $p>0$), or saturates to a plateau (if $p<0$). In most constructions $p\leq 1$, meaning that the parabola tends to flatten at large field values. The origin of this flattening is partially rooted in a rather generic feature of stringy inflationary models: the presence of multiple heavy moduli fields, corresponding to geometric features of the compactified extra-dimensions, which can be destabilized as the inflaton field is displaced far away from the minimum of its potential and backreact on the inflationary trajectory~\cite{Dong:2010in, McAllister:2014mpa}.\footnote{See also~\cite{Hebecker:2014kva, Bielleman:2016olv, Landete:2017amp} for flattening due to backreaction in concrete realizations of AMI which can be studied within supergravity.} Typically, these flattening effects kick in at $\phi\sim F\lesssim M_{p}$, possibly fitting nicely with general arguments against the validity of single-field EFT descriptions of large field inflation, such as the weak gravity conjecture~\cite{ArkaniHamed:2006dz} (see~\cite{Rudelius:2015xta, Brown:2015iha} for applications to axion inflation, and~\cite{Klaewer:2016kiy, Garg:2018reu, Ooguri:2018wrx} for related conjectures). Observationally, this feature is essential for the viability of \eqref{eq:monpot} as an inflationary potential. Indeed, power-like potentials with $p\geq 1$ are strongly constrained by CMB data, since they predict a large amplitude of primordial B-modes \cite{Akrami:2018odb}. Models with $p=1/3, 1/2$~\cite{McAllister:2014mpa} are still marginally compatible with CMB data. Here we will focus on $p<1$, while allowing also for concrete values of $p$ (such as 1/6) beyond the ones that have been obtained so far in concrete stringy setups. An explicit realization of AMI with $p<0$ --a possibility that we will consider-- has been provided in~\cite{Landete:2017amp}. 

Let us now discuss the second part of the potential \eqref{eq:monpot}, $V_{\rm cos}$. This contains the distinctive axionic {modulation}, superimposed on $V_{\text{mon}}$. Crucially, their amplitude $\Lambda(\phi)^4$ depends on the inflaton value.  We follow~\cite{McAllister:2014mpa} and parametrize this dependence as follows
\begin{equation}
\label{eq:amplitude}
\Lambda(\phi)^{4}=\Lambda_{0}^{4}\,e^{-\left(\frac{\phi}{\phi_{\Lambda}}\right)^{p_{\Lambda}}},
\end{equation}
with $\phi_{\Lambda}\gtrsim M_{p}$ and $p_{\Lambda}$ either positive or negative.
Putting \eqref{eq:monpot} and \eqref{eq:amplitude} together, we can write
\begin{equation}
\label{potential}
V(\phi)=m^{2}f^{2}\left[ \frac{1}{2p}\frac{F^{2}}{f^{2}}\left(-1+\left(1+\frac{\phi^{2}}{F^{2}}\right)^{p}\,\right)+\kappa e^{-\left(\frac{\phi}{\phi_{\Lambda}}\right)^{p_{\Lambda}}}\cos\left(\frac{\phi}{f}+\delta\right) \right]+V_0\,,
\end{equation}
where we have added a constant $V_0$, which ensures $V=0$ at the reheating minimum. 
The implications of $V_{\text{cos}}$ then depend on $p_{\Lambda}$, $p$ and the rescaled amplitude of the {modulation} $\kappa\equiv \Lambda_{0}^4/(m^{2}f^{2})$. 
Let us then first discuss separately the impact of $\kappa$, thereby initially neglecting the exponential prefactor. Close to $\phi=0$ we can then approximate \eqref{potential} by
\begin{equation}
\label{potapprox}
V(\phi)\approx m^{2}f^{2}\left[\frac{1}{2}\frac{\phi^{2}}{f^{2}}+\kappa\cos\left(\frac{\phi}{f}+\delta\right)\right].
\end{equation}
It is then straightforward to see that the potential \eqref{potapprox} exhibits local minima for  $\kappa \geq 1$, whereas for $\kappa < 1$ the oscillating part of the potential only gives rise to small bumps in the axion potential. In this work we are interested in local minima which appear close to the bottom of the inflationary potential (i.e.\ for $\phi/M_p\ll 10 $) and we will thus consider $\kappa \geq 1$.

Let us now return to the full potential. Depending on the sign of $p_\Lambda$ the amplitude of the  oscillations is exponentially suppressed or enhanced at large field values. The  value of $p_\Lambda$  is determined by  the  source of the non-perturbative effects that induce $V_{\rm  cos}$ and by moduli stabilization. See~\cite{Flauger:2014ana} for examples with both $p_\Lambda>0$ and $p_\Lambda<0$. We are interested in  $p_\Lambda>0$ since then oscillations are absent at $\phi\gg \phi_{\Lambda}$ and the flatness of the potential allows  to fit the CMB  without tunings, while still featuring local minima at smaller field values. This particular behavior of the inflationary potential is also somewhat similar to what has been used in the relaxion mechanism~\cite{Graham:2015cka}.

Finally, let us discuss the parameter $\delta$, which should be included on general grounds, since $V_{\text{mon}}$ and $V_{\text{cos}}$ have a priori no reason to be aligned. Furthermore, the choice $\delta=0$ leads to the presence of two degenerate minima at the bottom of the potential, which may lead to stable domain walls during the reheating phase, when the field can oscillate along the full potential. For these reasons, in what follows we take $\delta\sim 1$.

\begin{figure}[t]
\centering
\includegraphics[width=0.48\textwidth]{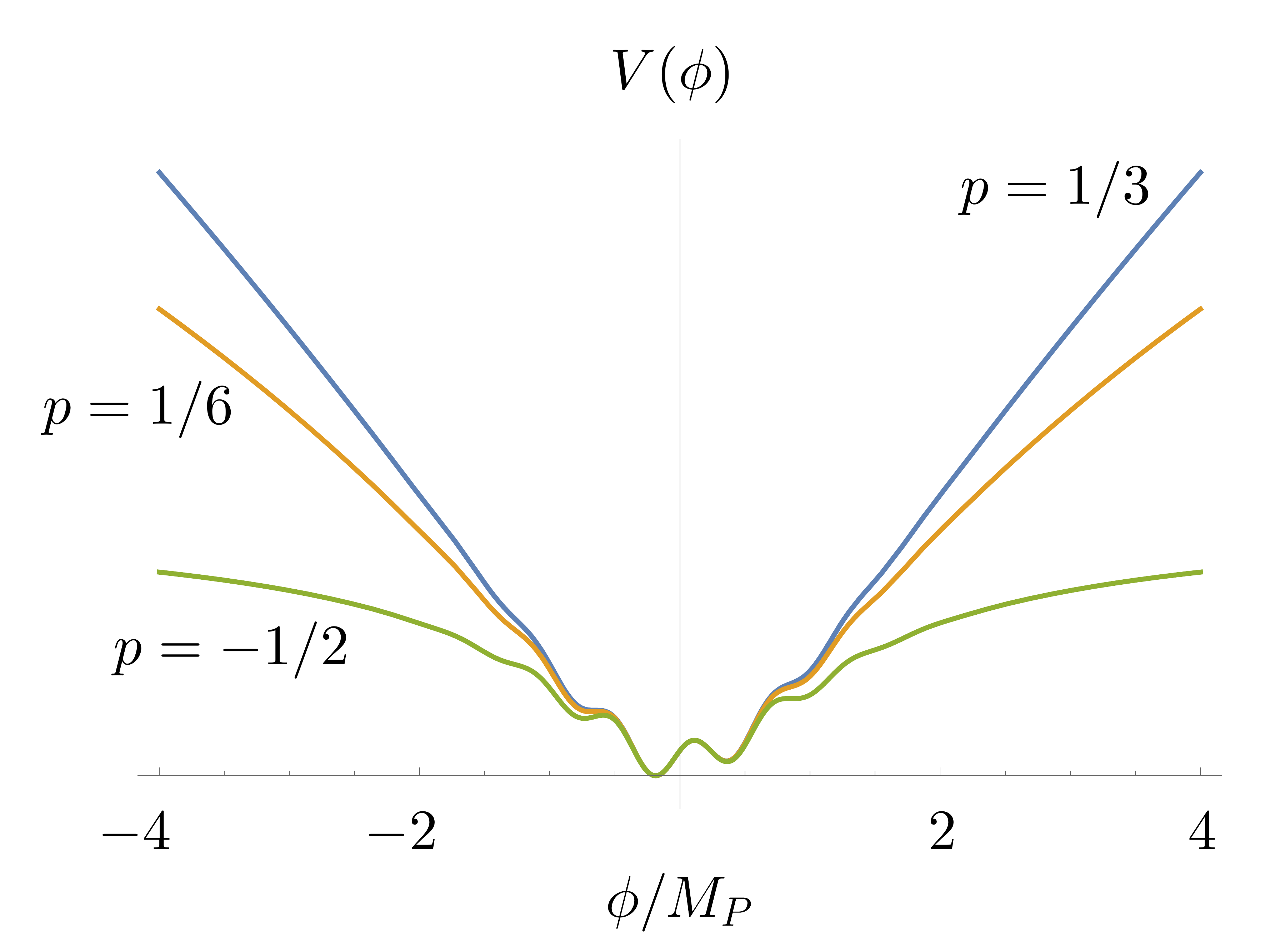}
\includegraphics[width=0.48\textwidth]{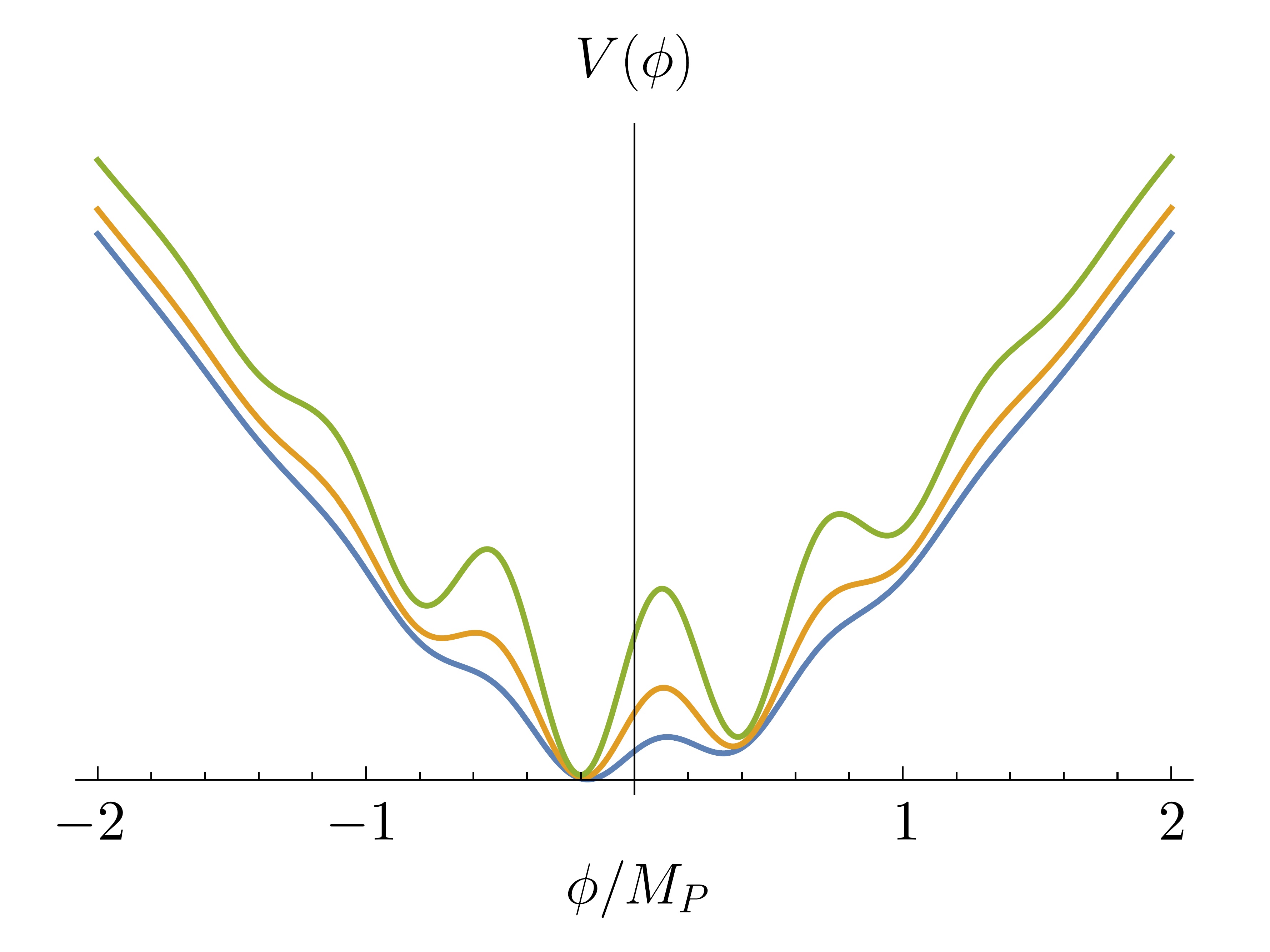}
\caption{\it Examples of the potential \eqref{potential}. Left: $\kappa=8$, $F=M_{p}$, $f=0.1 M_{p}$, $p_{\Lambda}=2$, $\phi_{\Lambda}=M_{p}$, $\delta=-1$ and $p=1/3, 1/6, -1/2$ from top to bottom. The parameter $p$ determines the asymptotic behavior of the potential. Right: $p=1/3$, $F=M_{p}$, $f=0.1 M_{p}$, $p_{\Lambda}=2$, $\phi_{\Lambda}=M_{p}$, $\delta=-1$ and $\kappa=20, 10, 5$ from top to bottom. Whereas the minima at small field values are affected by the value of $\kappa$, the inflationary trajectory at large field values is essentially unaltered varying this parameter alone. }
\label{fig:potential}
\end{figure}

The potential \eqref{potential} is shown in Figure \ref{fig:potential}, for $p=1/3,~1/6, -1/2$ from top to bottom, and with $p_{\Lambda}>0$. The figure illustrates the key feature of our inflationary potentials: beyond $\phi\sim 2 M_{p}$, the potential is essentially indistinguishable from a standard monomial, while at small field values the periodic axionic oscillations lead to a rich structure of local minima. 

Inflation along such potentials proceeds as follows. First, for large field values ($\phi\gg M_{p}$), the inflaton slowly rolls down the potential. This phase is the one responsible for the small CMB temperature anisotropies. Second, a regime of transient constant roll inflation (whereby $\ddot \phi \propto H\dot\phi$) can be achieved as the inflaton traverses one of the local minima at $\phi\sim M_p$. In this regime, super-horizon curvature fluctuations are exponentially enhanced, leading to PBH formation upon horizon re-entry (see Section \ref{results} for more details).
Interestingly, in our scenario the two phases (slow roll and constant roll) can be significantly decoupled from one another. The depth of the local minima is controlled by the parameter $\kappa$, which can be changed without affecting the inflationary potential in the region where CMB anisotropies are generated, as shown in Figure \ref{fig:potential}.

Due to the presence of local minima, the inflaton does not necessarily end up in the global minimum of the potential. In fact, in the regime $\kappa\gg 1$ the field typically gets classically stuck in one of the local minima closest to the global minimum. Let us estimate the tunneling rate to the global minimum from one of the nearest neighboring local minima. This is proportional to $e^{-S_{\text{t}}}$, where the tunneling action $S_{\rm t}$ can be easily estimated in the thin-wall approximation \cite{Coleman:1977py} as follows:
\begin{equation}
S_{\text{t}}\simeq \frac{27\pi^{2}}{2}\frac{\sigma^{4}}{(\Delta V)^{3}}\sim \frac{27\pi^2}{2}\frac{(\Delta\phi)^4(\kappa m^2 f^2)^2}{(m^2 f^2)^3}\sim\frac{27\pi^2}{2}\kappa^2\left(\frac{f}{m}\right)^{2}\,.
\end{equation}
Here we have approximated the tension of the bubble wall as $\sigma\sim\Delta\phi \sqrt{\Lambda_0^4}$, with $\Delta\phi\sim f$ the distance in field space between minima and $\Delta V$ the difference in height between the minima.
For $p=1$, the amplitude of the temperature anisotropies in the CMB implies $m \sim 10^{-6}M_{p}$. For $|p| < 1$, the CMB normalization depends also on $F$ and larger values of $m$ are allowed. Typical values of $f$ in string compactifications are $10^{-3} M_{p}\lesssim f \lesssim M_{p}$. Therefore, for $\kappa\sim\mathcal{O}(10)$ (as it will be in our  examples) $S_{t}$ is very large and tunneling to the global minimum is an extremely suppressed process, regardless of the prefactor.

The constant $V_{0}$ is chosen in such a way that the minimum where the inflaton stops has $V_0=0$.  After inflation ends, the inflaton then oscillates around an approximately quadratic local (or global) minimum and gives rise to the desired epoch of MD. Our Universe may then have a neighboring AdS vacuum (the global minimum of the full potential \eqref{eq:monpot}, if the inflaton gets stuck in a preceding local minimum). As shown above, this does not pose any cosmological threat to the stability of our Universe.

A potential (also inspired by AMI) with multiple approximate inflection points was considered in \cite{Ozsoy:2018flq} in the context of PBH formation (during radiation domination). It differs from ours in several respects that are worth mentioning. The most important difference is that the potential of~\cite{Ozsoy:2018flq} features potential oscillations also at large field values. This implies that the CMB is fit in this case by tuning very finely the parameters of two trigonometric functions, in such a way that the CMB scales coincide with a sufficiently flat region of the potential for large field values. The axion decay constant in the case of \cite{Ozsoy:2018flq} takes larger values:  $f\gtrsim 0.6~M_{p}$. In our case the CMB observables are essentially independent of the axion decay constant, which means $f$ takes somewhat smaller values: $f\gtrsim 0.2~M_{p}$. Another difference is that the examples of \cite{Ozsoy:2018flq} do not display multiple minima, but rather a successions of approximate plateaus. In that case inflation ends at the absolute minimum of the potential. Instead, we consider the possibility of several local minima where  the inflaton may get stuck. Finally, \cite{Ozsoy:2018flq}  focused on PBH formation during RD, while we take into account the possibility of a long MD epoch, which allows for a significant tuning reduction in the inflationary parameters. 

\section{Numerical examples}
\label{results}

In  this  section we compute the primordial power spectrum generated by the inflationary model presented above. We provide concrete numerical examples for which large PBH abundances with masses of interest for the DM problem are produced either during RD or an early phase of MD. We illustrate the advantage of MD over RD by quantitatively establishing the required amount of tuning of the parameters of the potential under both circumstances. 

Before doing so, we briefly review the mechanism by which the scalar primordial power spectrum can be enhanced in the presence of critical points in the inflationary potential. The dimensionless power spectrum \(\mathcal{P}_\mathcal{R}\) of the comoving curvature perturbation \(\mathcal{R}\) is defined as:
\begin{equation}
\label{powerspectrum}
\mathcal{P}_\mathcal{R}=\frac{k^3}{2\pi^2}|\mathcal{R}|^2.
\end{equation}
It is obtained from the Mukhanov-Sasaki equation:
\begin{align}
\label{curvaturemotion}
\mathcal{R}''+2\frac{z'}{z}\mathcal{R}'+k^2\mathcal{R}=0,\quad \text{where}\quad z^2 \propto a^2\epsilon\quad \text{and}\quad  \epsilon=-\frac{\dot{H}}{H^2}\,,
\end{align}
and the primes indicate derivatives with respect to conformal time. This equation is solved assuming the Bunch-Davies vacuum as the initial condition for the fluctuations at early times (i.e.\ when $k\gg z''/z\simeq aH$). The modulus $|\mathcal{R}|$ appearing in  \eq{powerspectrum} is evaluated at a later time such that $k\ll z''/z$. In the slow-roll approximation\footnote{See \cite{Liddle:1994dx} for its precise meaning or the appendix of \cite{Ballesteros:2014yva} for a brief summary.} $|\mathcal{R}|$ is very nearly constant for $k\ll z''/z$, leading to the usual expression
\begin{align} \label{srspectrum}
\mathcal{P}_\mathcal{R}=\frac{H^2}{8\pi^2\epsilon M_p^2}\,.
\end{align}
A small enough $\epsilon$ can thus lead to an enhancement of the primordial spectrum above its value at CMB scales. However, if the slow-roll regime ceases to be valid, even for a short period, the non-constant solution of $\mathcal{R}$ (which decays during slow-roll for $k\ll z''/z$) can become relevant and grow with time, thereby invalidating the approximation  \eq{srspectrum}. This is what happens in the examples we show below. The derivative of $\mathcal{R}$ with respect to the number of e-folds is, approximately:
\begin{equation}
\frac{d\mathcal{R}}{dN}\propto \exp\left(-\int \xi\, dN\right),
\end{equation}
where
\begin{align}
\label{activation}
\xi=3-\epsilon+\frac{\dot{\epsilon}}{H\epsilon}\,.
\end{align}
If $\xi$ becomes negative, which requires a large deviation from slow-roll, ${d\mathcal{R}}/{dN}$
grows with $N$ and the Mukhanov-Sasaki equation \eq{curvaturemotion} needs to be solved numerically for each mode $k$. An in-depth analytical description of the dynamics of the inflationary fluctuations for negative $\xi$ is given in \cite{Leach:2000yw, Leach:2001zf}; see also \cite{Ballesteros:2018wlw}.

As shown in \cite{Ballesteros:2017fsr} in the context of PBH formation during RD, $\xi$ can easily become negative if the inflaton encounters a sufficiently deep local minimum in its trajectory. Since 
$\epsilon\propto ({d\phi}/{dN})^2$, as the inflaton falls into the minimum, accelerating, $\epsilon$ grows and reaches a maximum value (which may or may not halt inflation temporarily). Then, as the inflaton climbs out of the potential well, its velocity quickly decreases, making $\epsilon$ hit very small values, provided that the minimum is deep enough. This sudden change of speed triggers a rapid growth of $|\eta|$ above 3, which is ultimately responsible for driving $\xi$ to negative values, making $\mathcal{R}$ increase exponentially in just a few e-folds. Clearly, the largest enhancement of the spectrum is obtained when the field is barely able to overshoot the local maximum after traversing through the local minimum. For PBHs produced during RD by this mechanism, $\epsilon$ needs to change by about 6 orders of magnitudes between CMB and PBH scales \cite{Ballesteros:2017fsr} in order to produce  $f_{\mathrm{PBH}}\sim 1$. In this case, $\mathcal{P_R}$ is enhanced by approximately 7 orders of magnitude between these scales and the slow-roll approximation can underestimate this  growth by several orders \cite{Ballesteros:2017fsr}. For PBH production by this mechanism during a phase of early MD, the change in $\epsilon$ and $\mathcal{P_R}$ need not  be as dramatic, since as we have seen in Section \ref{abundance} the power spectrum needs a significantly milder enhancement for PBH formation to occur.

We solve the Mukhanov-Sasaki equation (\ref{curvaturemotion}) numerically, together with the equation of motion for the background evolution of the inflaton field, to find  \(\mathcal{P}_\mathcal{R}\), defined as in (\ref{powerspectrum}). We refer the reader to \cite{Ballesteros:2017fsr} (see also  \cite{Chongchitnan:2006wx}) for the details of a practical numerical algorithm --that we have followed in this work-- through which the equation \eq{curvaturemotion} is solved using $N$ as time variable.  We are interested in parameter choices for the potentials such that the most recent constraints from Planck on the primordial spectrum are satisfied, see Appendix \ref{cosmoparams}.

The potential \eq{potential} can feature many local minima whose depth grows as the inflaton travels from larger to smaller values.  During its trajectory the inflaton passes through several of these minima, and the enhancement of the primordial spectrum becomes most pronounced when it goes through the next-to-last minimum, before stopping definitively; since it is in this region that it slows down the most. The depth of the minima is controlled by the parameter $\kappa$,  as described in Section \ref{sec:potential}. Any large enough value of $\kappa$ ensures the existence of minima which may lead to abundant PBH formation. The actual value of the abundance is determined by the speed of the inflaton as it climbs out of the next-to-last minimum before reheating. This speed is, in turn, fixed by the precise value of $\kappa$. In our examples, we adjust the parameter $\kappa$ to obtain $f_{\rm PBH}\sim\mathcal{O}(1)$, and find $\kappa\sim\mathcal{O}(10)$. The smaller amplitude of $\mathcal{P}_\mathcal{R}$ required to account for all DM with PBHs formed from MD tends to reduce the number of e-folds that the inflaton field spends traversing the local minimum responsible for PBH formation with respect to the case of RD. Therefore, imposing $f_{\rm PBH}\sim\mathcal{O}(1)$ with masses in the window \eq{win}, some examples of potentials that are ruled out for PBH formation during RD due to an excess of inflation --see \cite{Liddle:2003as} and  equation \eq{totalefolds}-- may become viable changing $\kappa$ appropriately if the PBHs form during an early MD phase. The same can be expected to occur for other models with an approximate inflection point. 

\begin{figure}[t]
\centering
\includegraphics[width=0.58\textwidth]{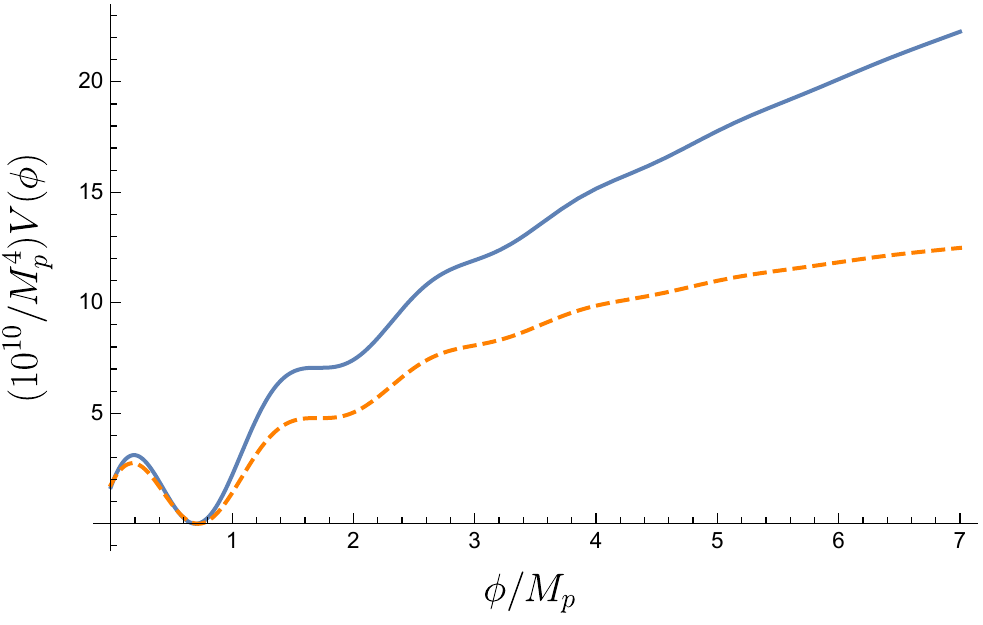}
\caption{\it Potential \eq{potential} for the parameters in Table \ref{tab:parameters}. The solid blue curve corresponds to example 1 (MD) and the dashed orange curve to example 2 (MD).}\vspace{0.3cm}
\label{fig:potentialsexamples}
\end{figure}
 
Let us now discuss the effects of the rest of the parameters of the potential. We start with $F$ and $\phi_{\Lambda}$, which control the location in field space at which the flattening effects kick. We will take them to be of order $M_{p}$, in agreement with the expectations discussed in Section \ref{sec:potential}.

The parameter \(f \lesssim M_p\)  governs the width of the local minima. We can distinguish between different scenarios depending on its value. For values of $f$ close to $M_p$, the inflaton encounters at most one local minimum before inflation ends. In this limit the model is essentially an implementation of the standard mechanism of PBH production from a quasi-inflection point. We will not consider this situation here, but we remark that it is a possibility capable of producing an interesting population of PBHs for the DM problem, if $\kappa\sim \mathcal{O}(1)$. In the opposite limit,  for $f\ll 1$, the inflaton may roll all the way down to the global minimum and oscillate in a region of the potential which can encompass several local minima. This case is not relevant for PBH formation and we will not consider it either.\footnote{However, it may present interesting consequences for reheating~\cite{Hebecker:2016vbl}.}
We thus focus on intermediate values of $f$ for which the inflaton still travels over several minima before inflation ends. We find that if \(f\lesssim0.1 M_p\) and $\mathcal{P_R}\sim\mathcal{O}(10^{-4})$ is imposed at its maximum, the field generically does not spend enough time (typically at most $\sim5$ e-folds) on the local minimum previous to the end of inflation. In this case inflation does not last long enough to solve the horizon problem.\footnote{Approximately 15 e-folds are required between the scale at which perturbations are enhanced (for the mass of the PBHs to fall in the window \eq{win}) and the end of inflation.} We focus on $f=0.2 M_p$ in our examples and return to the case $f\lesssim 0.1 M_p$ at  the end of the  section. 

We will consider two concrete choices for the parameter $p$, one with $p>0$ (shown in Figure \ref{fig:potentialsexamples} as the blue, solid line), and the other one with $p<0$ (orange, dashed). In particular, we find that the largest positive value of $p$ which is compatible with the latest Planck data and at the same time leads to a significant amount of light PBHs is smaller than $p=1/3$, which is the smallest positive exponent for which an explicit string construction currently exists \cite{McAllister:2014mpa}. We choose $p=1/6$ to produce our first example and $p=-1/2$ for our second example. We focus on the case in which the inflaton gets stuck in the local minimum closest to the global minimum, which corresponds to rather large values of $\kappa$. The scenario with the inflaton rolling all the way down to the global minimum can also be easily realized for both $p>0$ and $p<0$, by taking smaller values of $\kappa$. In both cases, we set $V=0$ at the minimum of the potential where the field ends its trajectory (see Figure \ref{fig:potentialsexamples}).

\begin{table}[t]
\renewcommand{\arraystretch}{1.25}
\centering
\resizebox{\textwidth}{!}{
\begin{tabular}{|c|c|c|c|c|c|c|c|c|c|} 
\hline 
& $f$ & $\delta$ & $p_\Lambda $ & $\phi_\Lambda$ & $p$ & $F$ & $m\cdot 10^{5}$ & $V_0\cdot 10^{11}$ & $\kappa$ (MD/RD) \\[2pt] \hline
{\color{blue}$\bullet$} Example 1 & $0.2~M_p$ & -1 & 1 & $1.15M_p$ & 1/6 & $0.75~M_p$ & $3~M_p$ & $-5.1~M_p^4$ & 8.254/8.254538\\ \hline
{\color{orange}$\bullet$} Example 2 & $0.2~M_p$ & -1 & 1 & $1.15M_p$ & -1/2 & $1.85~M_p$ & $2~M_p$ & $2.0~M_p^4$ & 14.986/14.986471\\ \hline
\end{tabular}}
\vspace{0.2cm}\\
\resizebox{\textwidth}{!}{
\begin{tabular}{|c|c|c|c|c|c|c|c|c|c|}
\hline
& $\phi_0$ & $n_s$ & $r$ & $\alpha$ & $N_0$(MD/RD) & $f_\mathrm{PBH}$(MD/RD) & $M_\mathrm{PBH}$\\[2pt] \hline
{\color{blue}$\bullet$} Example 1 & $6.85~M_p$ & 0.970 & 0.068 & -0.009 & 51/57 & 0.9/0.4 & $10^{-13}~M_\odot$\\ \hline
{\color{orange}$\bullet$} Example 2 & $6.04~M_p$ & 0.970 & 0.036 & 0.02 & 52/56 & 0.2/0.6 & $10^{-15}~M_\odot$\\ \hline
\end{tabular}}
\caption{\it Parameters and predictions for the two potentials in Figure \ref{fig:potentialsexamples}. The parameters give the correct normalization of the spectra at  $k=0.05$ {\rm Mpc}$^{-1}$, corresponding  to $\phi=\phi_0$. The CMB parameters (see Appendix \ref{cosmoparams}) are given at this scale, and $N_0$ denotes the number of e-folds from $\phi_0$ to the end of inflation. The values of $\kappa$ are given  with  the precision needed to attain the corresponding $f_{\rm PBH}$ in each case. We have set \(T_m=10^4\,\rm{GeV}\) and $\mathcal{I}=3.1$ for Example 1, and \(T_m=2\cdot10^5\,\rm{GeV}\) and $\mathcal{I}=4.4$ for Example 2.}
\label{tab:parameters}
\end{table}

\begin{figure}[p]
\centering
\includegraphics[width=0.45\textwidth]{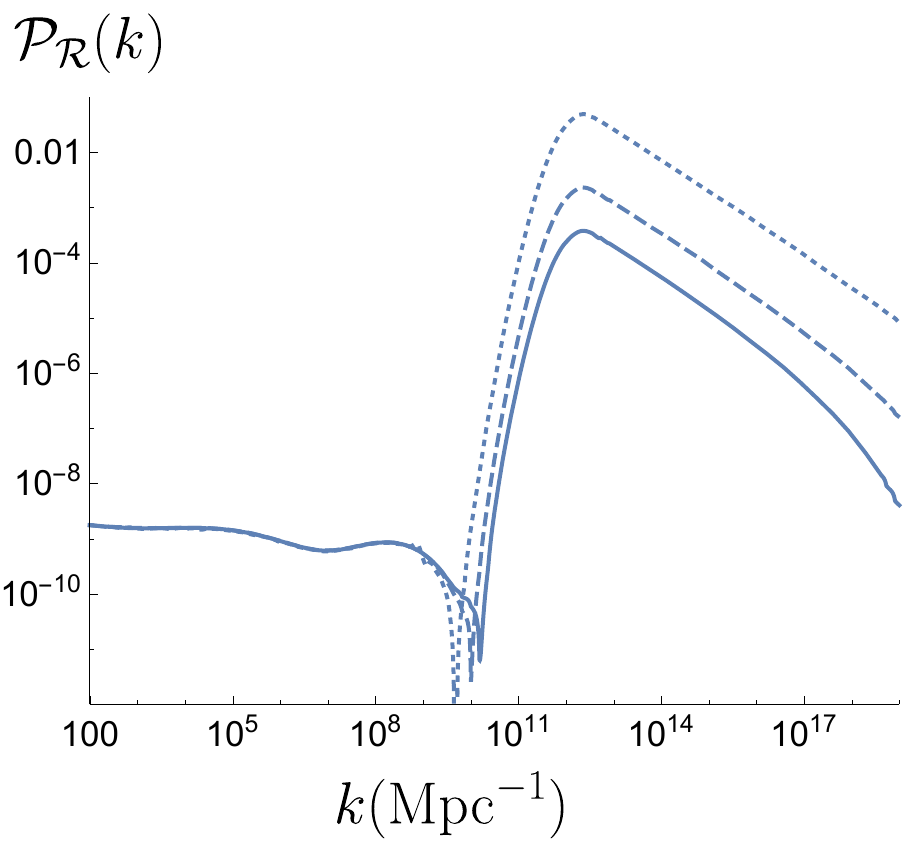}
\includegraphics[width=0.45\textwidth]{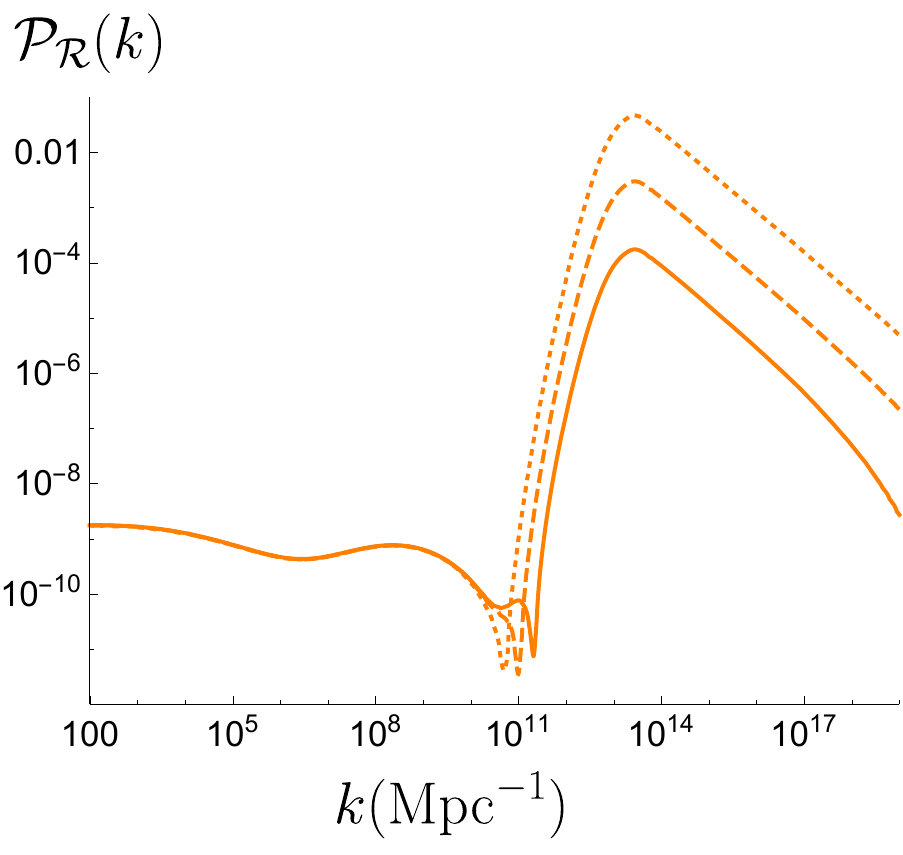}
\caption{\it Primordial spectra for Examples 1 (left) and 2 (right). The dotted and dashed lines correspond to the parameters choices  in Table \ref{tab:parameters}  labeled RD and MD, respectively. The solid lines (not included on the table) are shown for ease of comparison and correspond to setting $\mathcal{I}=1$, together with $\kappa=8.253$ for Example 1, and $\kappa=14.984$ for Example 2.}\vspace{0.2cm}
\label{fig:example2}
\end{figure}

\begin{figure}[t]
\centering
\includegraphics[width=.39\textwidth]{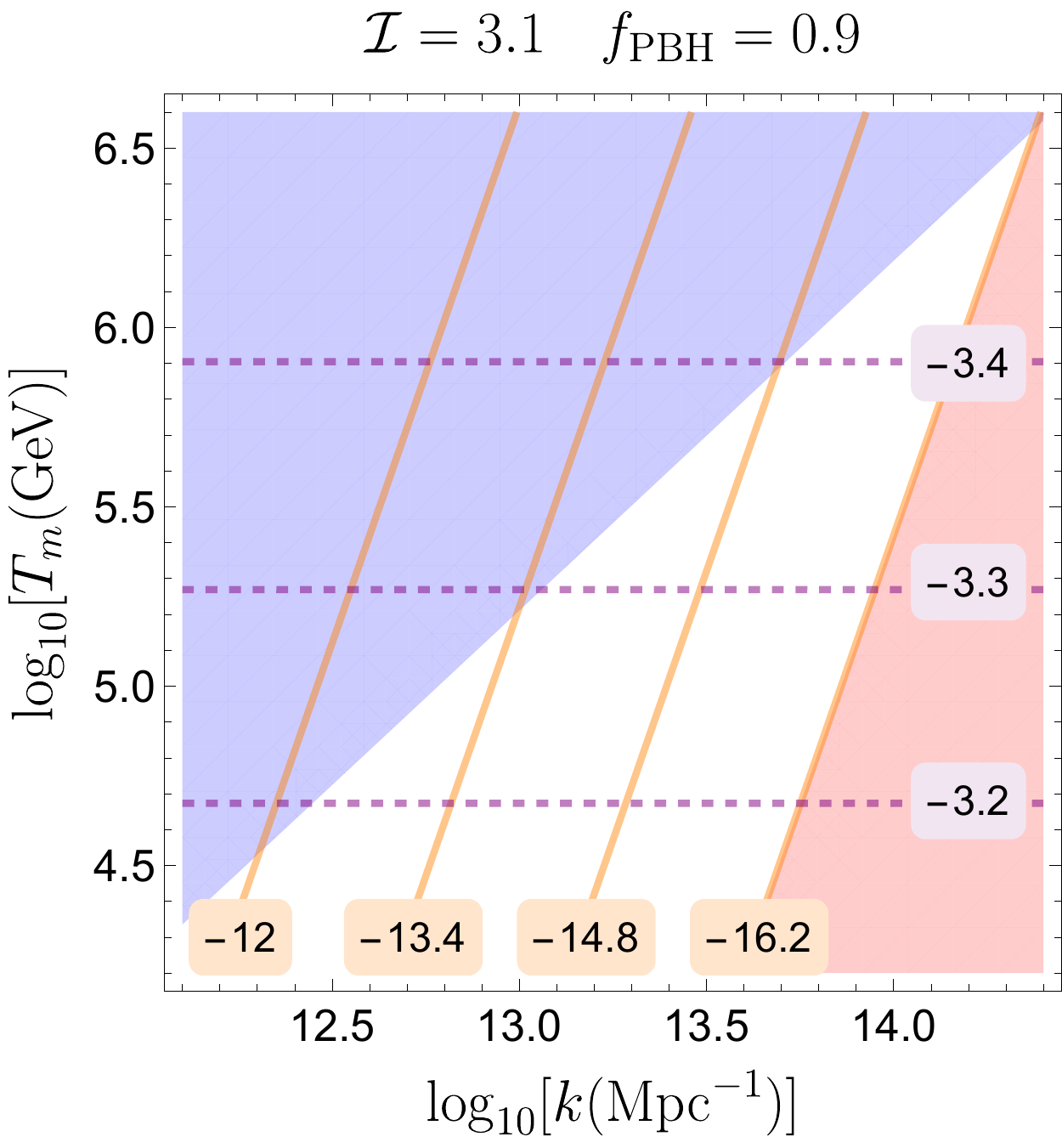}\hfill
\includegraphics[width=.61\textwidth]{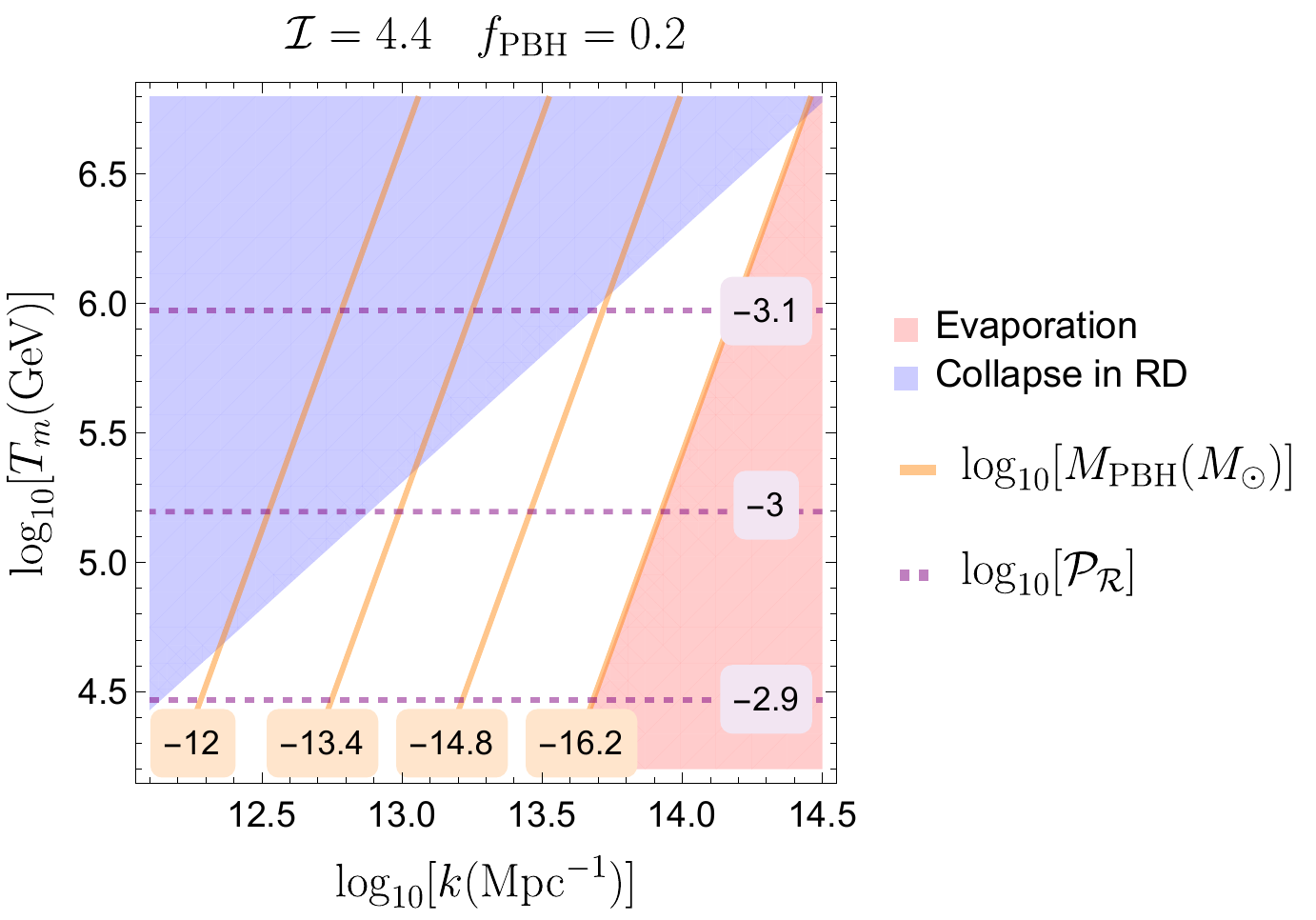}
\caption{\it Effect of changing the parameter $\mathcal{I}$ in (\ref{betaam}). Legend and colors as in Figure \ref{abundancemd}. The values of $\mathcal{I}$ and $f_\mathrm{PBH}$ depicted correspond to examples 1 (left) and 2 (right).}
\label{abundanceChangeI}
\end{figure}

{In Figure \ref{fig:example2} we show examples of primordial spectra  for the potential \eq{potential} with $p=1/6$ (left panel, {\it Example 1}) and $p=-1/2$ (right panel, {\it Example 2}). For each choice of $p$ we show three different spectra, obtained by varying only the value of the parameter $\kappa$, which controls their height. The values of $\kappa$ corresponding to the dotted and dashed spectra are given in Table \ref{tab:parameters}, indicated as RD and MD, respectively. The dotted spectra, which have a higher peak, originating from a more severe tuning of $\kappa$, provide a significant fraction of the DM ($f_{\rm PBH}\sim 0.1 - 1$) for PBH formation during RD. Indeed, the values of $f_{\rm PBH}$ indicated in Table \ref{tab:parameters} have been computed assuming formation during RD for the dotted spectra and during MD for the dashed spectra.\footnote{{The abundance in the case of RD has been computed using equation \eq{betard}, which assumes Gaussian fluctuations. Due to the large change in the inflaton velocity in the region of the potential responsible for PBH formation, non-Gaussianities are produced. They can be estimated from the value of the second slow-roll parameter on its minimum, see e.g.\ \cite{Atal:2018neu}. For the examples presented in Section~\ref{results}, we obtain $f_{\rm NL}\simeq 0.4$, in line with the models discussed in \cite{Atal:2018neu}. A full calculation of the effect of non-Gaussianities in models with an inflection point is not available yet and goes beyond the scope of our work. In reference \cite{Franciolini:2018vbk} it is estimated that non-Gaussianities amount to a correction of order $60\%$ to the exponent of equation  \eq{betard}, from a partial cumulant  resummation. This is a large effect --which goes in the direction of reducing $\beta$ (for fixed $\sigma$ and $\delta_c$)-- and reflects the  exponential sensitivity  of the abundance in the RD case. A change of the exponent of equation  \eq{betard} can be compensated by a modification of $\sigma$ by the same  amount,  which can  be easily arranged modifying slightly the parameters of the potential. Similarly, in RD a change in $\delta_c$ (whose precise value is uncertain) can  be  compensated analogously. The effect of non-Gaussianities on PBH formation during MD in models of inflation with an inflection point remains to be studied.}}  Table \ref{tab:parameters} contains also the predictions for the inflationary observables and the peak PBH masses. The abundances for the dashed spectra (assuming formation during MD and  $T_{m}\lesssim 10^{6}~\text{GeV}$) have been computed using \eqref{betaam} with \(f_q(q_c)=1\) (following \cite{Harada:2017fjm}), which takes the effect of angular momentum into account. For the examples summarized in  Table \ref{tab:parameters} we have chosen $\mathcal{I}=3.1$ for Example 1, and $\mathcal{I}=4.4$ for Example 2. These choices ensure that the resulting PBH mass distribution evades the evaporation constraints of \cite{Carr:2009jm} (see also \cite{Arbey:2019vqx,Ballesteros:2019exr}) for non-rotating PBHs, as can be appreciated in Figure \ref{abundanceChangeI} (which can be compared with Figure \ref{abundancemd}).}

\begin{figure}[p]
\centering
\includegraphics[width=0.45\textwidth]{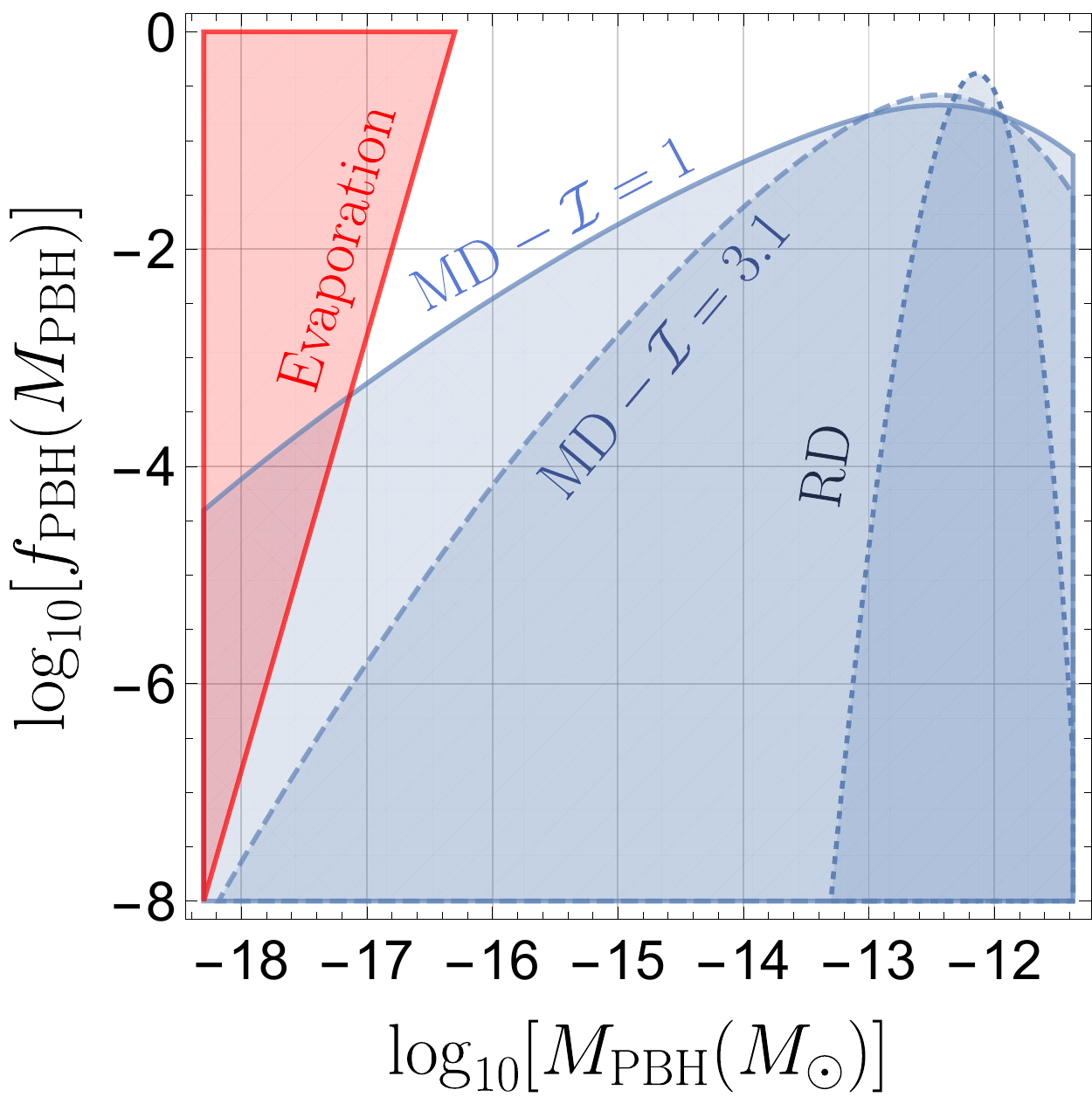}
\includegraphics[width=0.46\textwidth]{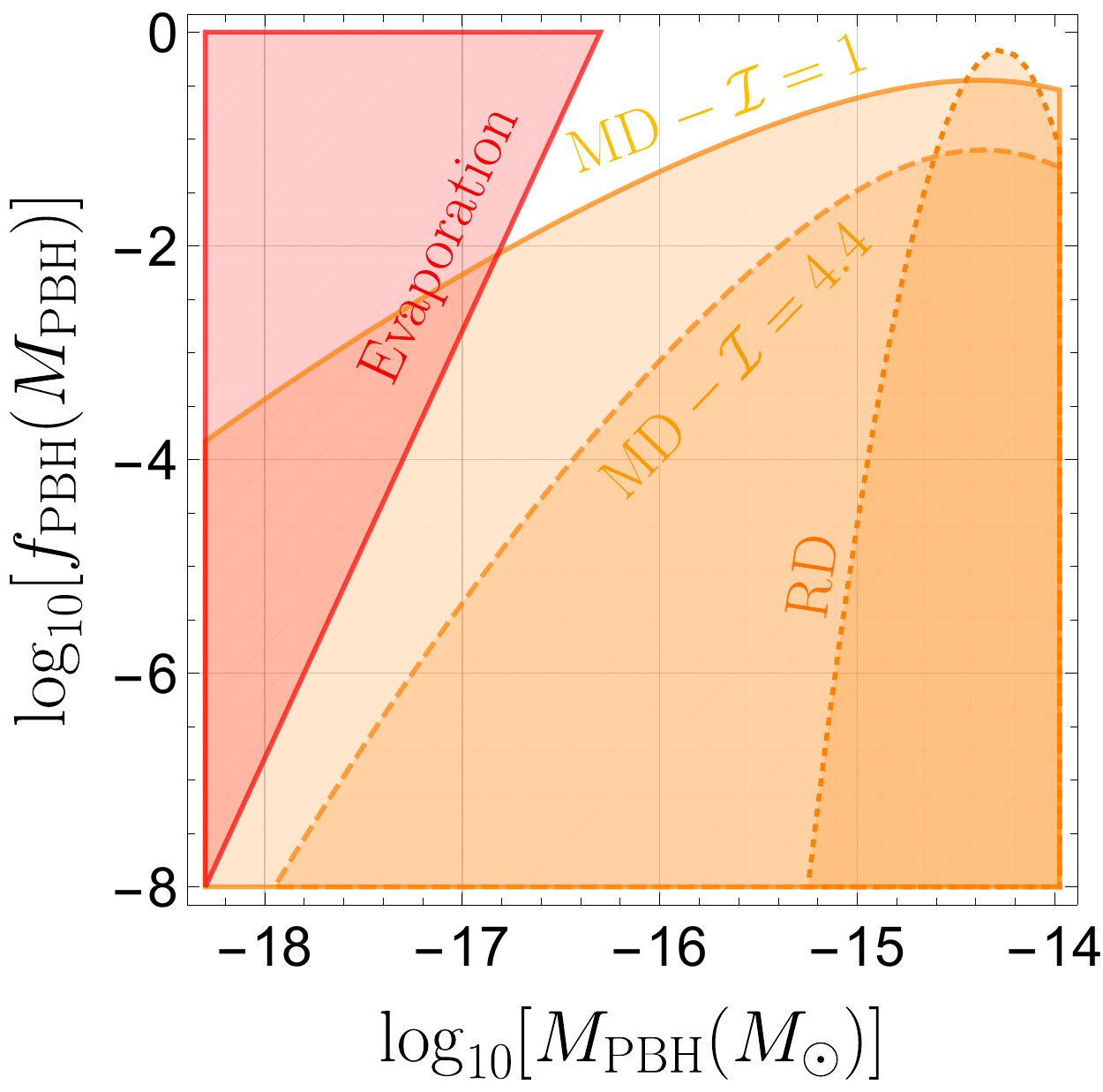}
\caption{\it Distribution of PBH masses for Examples 1 (left) and 2 (right). The largest masses shown in each panel correspond to the last modes that undergo collapse during the MD era. The red region corresponds to the bound in \cite{Carr:2009jm}, see also \cite{Arbey:2019vqx,Ballesteros:2019exr}. Although stronger bounds exist (see e.g.\ \cite{Arbey:2019vqx} for the extragalactic gamma-ray background applied to spinning black holes  and \cite{Laha:2020ivk} for the Galactic one), these can also be evaded by increasing $\mathcal{I}$. We have also included for comparison the curves obtained by setting $\mathcal{I}=1$, with $\kappa=8.253$ for Example 1, and $\kappa=14.984$ for Example 2. The largest masses shown in each panel correspond to the last modes that undergo collapse during the MD era (in the MD scenario). The shape of the distribution for larger masses depends (unless one assumes pure RD right after inflation) on the details of the transition between MD and RD, but we expect a rapid decay of the abundance. 
}\vspace{0.2cm}
\label{fig:example3}
\end{figure}

{Since the suppression of the PBH formation probability away from the peak of the spectrum is much milder in MD than in RD, the mass distribution functions (depicted in Figure \ref{fig:example3}) decay much more quickly in the latter case. This means that the PBH evaporation bounds become more difficult to evade (for a given $k$ value at the peak of the spectrum) if the PBHs form during an early MD era. This is a generic challenge for any scenario of light PBH formation during MD. As we just mentioned, we find that the extragalactic gamma-ray evaporation bounds can be evaded in our examples, provided that the parameter $\mathcal{I}$ in (\ref{betaam}) is around $\mathcal{I}\sim3$. Slightly larger values of $\mathcal{I}$ would also allow to circumvent the (more stringent) constraint from Galactic emission of \cite{Laha:2020ivk}, as well as the rest of the evaporation bounds (including those for spinning PBHs). Whether such values of $\mathcal{I}$ are feasible depends on the shape of the power spectrum as well as on the details of the collapse \cite{Harada:2017fjm}. In principle, $\mathcal{I}$ can be computed  --borrowing the formalism of \cite{1969ApJ...155..393P} for galaxy formation-- given the primordial power spectrum \cite{Harada:2017fjm}. For primordial spectra of the form $\mathcal{P_R}=k^{-y}$, with $1>y>0$, it was found in \cite{1969ApJ...155..393P} that values of $\mathcal{I}>1$ correspond to $y \lesssim 1/2$. Coincidentally, the slopes of the spectra in Figure \ref{fig:example2} are close to $y=1/2$ for  $k$ larger than the peak location, which corresponds to $\mathcal{I}\simeq 1.1$ \cite{1969ApJ...155..393P}. 

The modes with $k$ to the right of the peak in Figure \ref{fig:example2} are the ones most relevant for the evaporation bounds (see Figure \ref{abundanceChangeI}), since the PBH mass goes as $\sim k^{-3}$ and longer modes reenter the horizon after shorter ones. Moreover, if we assume that the collapse happens quickly enough, only the modes that are already inside any given Hubble patch can affect the collapse of that patch, and the spectra of Figure \ref{fig:example2} may indeed be well approximated by $\mathcal{P_R}=k^{-y}$ in the region relevant for computing the value of $\mathcal{I}$ for masses lighter than the peak mass. Taken at face value, this argument and the aforementioned results of \cite{1969ApJ...155..393P} exclude $\mathcal{I}\sim 3$ for the spectra of Figure \ref{fig:example2}, which would make them incompatible with the less stringent evaporation bounds of \cite{Carr:2009jm, Arbey:2019vqx,Ballesteros:2019exr}. The mass distributions depicted with continuous lines in Figure \ref{abundanceChangeI} illustrate this point.\footnote{{These mass distributions correspond to the spectra drawn with continuous lines in Figure \ref{fig:example2}. The potentials giving these spectra have values of $\kappa$ different from those in Table \ref{tab:parameters}, as indicated in the caption of Figure \ref{fig:example2}. These modified values of $\kappa$ have been chosen to maintain $f_{\rm PBH}\sim \mathcal{O}(0.1 - 1)$ despite the reduction in $\mathcal{I}$ from $\sim 3$ to 1, see equation  \eq{betaam}.}} Indeed, for $\mathcal{I}=1$ (see Figure \ref{abundancemd}) only a primordial power spectrum that decreases sufficiently rapidly at large $k$ can take full advantage of the MD phase without clashing with the evaporation bounds.}

{Whereas the previous discussion illustrates the challenge that evaporation bounds represent for abundant PBH DM in the case of PBH formation during MD, one should not conclude from it that this scenario should be discarded. There are several arguments which suggest that computing the  abundance with \eq{betaam} and assuming $\mathcal{I}\simeq 1$ (instead of a slightly larger value) is just but one possible approximation to the actual result. For instance, as pointed out in \cite{Harada:2017fjm}, first order contributions to the angular momentum of the collapsing overdensity could be more relevant than the second order contributions, depending on the effects of the duration of the collapse on the angular momentum. If this is the case, the collapse fraction can be severely suppressed and becomes quantitatively similar to the case of collapse during radiation domination. Furthermore, higher order corrections to \eqref{betaam} may also suppress the collapse fraction, and while it is reasonable to assume that perturbatively $f_c(q_c)\sim 1$ (see equation \eqref{betaam}), a non-perturbative analysis (which could make this fraction significantly smaller) is currently lacking \cite{Harada:2017fjm}. It is thus conceivable that the realistic collapse fraction depends on $k$ (and thus on the PBH mass), in such a way that the expression \eqref{betaam} (for $\mathcal{I}\simeq 1$) would apply reasonably well to scales close to the peak of the power spectrum, while the fraction is more suppressed away from the peak. In view of this, we regard the quantity $\mathcal{I}\sim \mathcal{O}(1)$ as a proxy to quantify the degree of suppression of the mass function away from its peak, allowing to quantify the theoretical uncertainties on \eq{betaam}.}

{Therefore, while PBH formation during MD is less suppressed than during RD --with the PBH abundance having an approximately monomial dependence on the variance of the density fluctuations rather than an exponential one--, the details of a possible further suppression due to the effects of angular momentum (and, importantly, their implication for evaporation bounds) require a dedicated (and possibly numerical) analysis, which goes beyond the scope of the present work.}

From our examples, we are able to extract the amount of tuning on the parameter $\kappa$ which is required in order to obtain $f_{\mathrm{PBH}}\sim 1$ in our setup with PBH formation during MD and RD. We do so defining\footnote{This way of quantifying the tuning was also used in \cite{Hertzberg:2017dkh}.}
\begin{align}
{\rm tuning}\coloneqq\bigg{|}\frac{\Delta \kappa}{\kappa}\bigg{|}\,,
\end{align}
where  $\Delta \kappa$ is the difference in $\kappa$ between a successful example with $f_{\rm PBH}\sim 1$ and the closest $\kappa$ which invalidates the example (typically by producing $f_{\rm PBH}$ away from 1), with the rest of the parameters kept fixed. In other words, $\Delta\kappa$ is given by the minimal precision with which $\kappa$ needs to be specified to obtain $f_{\rm PBH}\sim 1$. In the case of MD, we find that $\kappa$ has to be chosen with a relative precision (tuning) of order $10^{-2}~\%$ for $p=1/6$ and $6\cdot 10^{-3}~\%$ for $p=-1/2$. For RD, we instead find that the required tuning is increased to order $10^{-5}~\%$ for $p=1/6$ and $6\cdot 10^{-6}~\%$ for $p=-1/2$. Thus, an early phase of MD alleviates the tuning of the inflationary parameters in our setup by three orders of magnitude. {We note that this suppression is essentially independent of the value of the parameter $\mathcal{I}\sim\mathcal{O}(1)$.} While we have obtained our results for a specific inflationary model, our conclusions on the tuning are expected to remain valid for other models, since they are mainly a consequence of the model-independent results of Section \ref{abundance}.

Let us now comment on  the case $f\lesssim 0.1~M_{p}$. As we mentioned earlier, in this case the model is not able to provide enough e-folds of inflation to solve the horizon problem. Nonetheless, this scenario possesses an interesting feature: very large negative values of $\xi$ (defined in equation \eqref{activation}) can be achieved. Thus, according to \cite{Atal:2019cdz}, if the PBHs form during RD, then another mechanism of PBH formation, different from the one we have been considering so far, can become important. Large non-Gaussian fluctuations may prevent the field from overshooting the local maximum, implying that some regions of the Universe remain stuck in the local minimum while inflation takes place. The vacuum bubbles that form are then able to collapse  after inflation and end up in the form of PBHs. We will not delve further into this case, but we briefly mention one possibility to make this region of parameter space viable: if the inflaton could get stuck in one of the local minima during some e-folds before tunneling into the global one, then an additional phase of false vacuum inflation could be achieved, yielding the necessary amount of inflation to solve the horizon problem. As we have shown in Section \ref{sec:potential}, this scenario is extremely unlikely in our setup because the tunneling action is huge. However, it may be possible to envisage mechanisms, possibly involving extra fields,  by which the barrier between the local and global minimum can be lowered after the field gets stuck in the former, and thus increase the tunneling rate. Scenarios of these sort have been considered for the graceful exit problem of inflation in a false vacuum, see e.g.~\cite{DiMarco:2005zn}.

\section{Conclusions}
\label{sec:conclusions}

PBHs are intriguing but theoretically pricey DM candidates. In the context of canonical single-field inflation, the standard mechanism to produce them relies on an  approximate inflection point in the potential, which generates large primordial fluctuations at small distance scales. Regions where the associated density fluctuations are large enough collapse into PBHs when these perturbations become sub-horizon during the radiation epoch. The presence of the inflection point is generically engineered for the sole purpose of PBH  production. The abundance of the PBHs that are generated during RD is exponentially sensitive to the amplitude of the primordial fluctuations at those scales, which is controlled by the detailed properties of the inflection point. For this reason, in order to account for a large fraction of the DM, one or more parameters of the potential need to be adjusted with high precision. The {\it raison d'être} for inflation is the solution of outstanding cosmological problems which can be phrased in terms of tuning. It is therefore somewhat disappointing that the inflationary potential needs to be tweaked to address the DM problem as well.

These concerns about PBH  DM are specific of models of canonical single-field inflation. However, so far they have not been circumvented  by considering non-canonical inflation or models of inflation with several fields, where analogous issues arise. Other scenarios of PBH formation (not based on inflation) also require tuning in one way or another. 

In this paper, we have presented a scenario of PBH DM which partially alleviates these issues. The scenario is based on two ingredients: an axion-like inflationary potential with oscillations superimposed on a non-periodic term and a long phase of MD after inflation. We have focused on the case in which this phase is due to the oscillations of the inflaton around an approximately quadratic minimum of the potential. In contrast to most models of PBH formation from canonical single-field inflation, the existence of critical points in our potential is a generic property that arises when axionic `wiggles' are large enough compared to the non-periodic part of the inflaton potential. Our setup is inspired by a popular realization of large field inflation in compactifications of string theory, known as axion monodromy inflation. This class of models is generally characterized by a flattening of the inflationary potential at large, transplanckian, field values. We take advantage of this feature in our setup and work with inflationary potentials which present two distinct regions. The first one, located at large field values, is flat enough to allow for a standard phase of slow-roll inflation in which fluctuations with the right properties to fit the CMB are generated. The second region is close to the global minimum of the potential and can exhibit several local minima which ultimately lead to PBH formation. Although abundant PBH formation is not a completely general feature of this class of potentials, the appearance of multiple minima reduces significantly the level of engineering that is needed for the presence of an adequate approximate inflection point. Identifying the inflaton with a modulus field --which couples only gravitationally to light degrees of freedom-- the second ingredient of our scenario can be realized: the decay rate of the inflaton is Planck-suppressed, preheating can be avoided and a long epoch of MD after inflation can take place. During this period, the Universe is approximately pressureless. This results in an enhanced fraction of PBHs, even when angular momentum effects of the collapsing regions are taken into account. 

We find that reheating temperatures around $10^{5}~\text{GeV}$ -- $10^{6}~\text{GeV}$ are particularly advantageous for PBH DM from an early phase of MD. For such temperatures, the size of the primordial power spectrum at scales $k\sim 10^{13}~\text{Mpc}^{-1}$ -- $10^{14}~\text{Mpc}^{-1}$ that is required to obtain an $\mathcal{O}(0.1-1)$ fraction of the DM in PBHs with observationally viable masses ($M_{\text{PBH}}\sim 10^{-16}M_{\odot}$-- $10^{-14}M_{\odot}$) is of the order $\mathcal{P_R}\sim10^{-5}-10^{-3}$. This contrasts with the $\mathcal{O}(10^{-2})$ primordial spectrum that is needed {(under the assumption of Gaussian fluctuations)}  to attain the same abundance if PBHs are formed during RD.

None of the single-field scenarios we are aware of achieves a lower level of tuning than the one presented here. Our examples show that a long epoch of early MD translates into a relaxation of three orders of magnitude on the tuning of the single most relevant parameter for the specific class of potentials we consider, which controls the depth of the minima.  This occurs for potentials which grow as a power-law as well as those that saturate to a plateau at large field values, both of  which can accommodate the necessary enhancement of curvature fluctuations at small scales and  an amount of inflation consistent with the aforementioned reheating temperatures. Interestingly, our scenario features a scalar spectral index in excellent agreement with the CMB, which is not easy to achieve in canonical single-field models of inflation of PBH DM. Also, our examples favor an axion decay constant around $0.2M_{p}$, in agreement with general arguments against axion periodicities larger than $M_p$. 

{The slower fall-off of the PBH abundance for PBHs formed during MD --as compared to the standard scenario of RD-- implies that a larger amount of PBHs with masses smaller than the peak mass of the  distribution can be produced. Since the window where PBHs can comprise the totality of the DM is bounded from below due to Hawking evaporation (with concrete limits depending on the properties of the black holes and the observations that are considered), any scenario of PBH DM with PBHs originating in an early phase of MD is more susceptible to be strongly constrained from these bounds than in the RD case. Whether the bounds can be avoided, depends on the role of angular momentum in the PBH formation process (as it affects the abundance) and the spin of the resulting PBHs (which enhances the evaporation rate). For the inflationary potential that we propose, we have found that the bounds can be avoided provided that an $\mathcal{O}(1)$ parameter related to the angular momentum of the collapsing overdensity takes a sufficiently large value (effectively narrowing the width of the mass distribution). Although there exists an approximate formalism that allows to compute this parameter from a given power spectrum, further work is needed to understand the compatibility of PBH DM in this scenario with different evaporation bounds.

{Exploring  the possibility of primordial spectra with very narrow peaks from concrete models of inflation might be an interesting line to pursue in this context, as this could allow for narrower mass distributions, potentially less dangerous for current and future evaporation limits. Besides, in our opinion, the idea of PBH production from inflationary potentials with several minima deserves further study. Finally, possible and necessary extensions of our work are} the  inclusion of non-Gaussianities and quantum diffusion, which may have a quantitative effect in the required level of tuning.

Gravitational collapse during and early phase of MD for PBH DM has not been studied (in particular recently) as much as the case of RD. This is probably due to the relative simplicity of the Press-Schechter formalism which has been extensively is used} in the latter case. Given its advantageous more  moderate sensitivity to the variance of the density fluctuations, a better characterization of the process from numerical analyses in MD would undoubtedly contribute to set this scenario on a more balanced footing with respect to the RD case.

\section*{Acknowledgments}

We thank Vicente Atal,  Jaume Garriga, Shao-Jiang Wang and Alexander Westphal for discussions. We thank Tomohiro Harada, Chul-Moon Yoo, Kazunori Kohri and Ken-Ichi Nakao for correspondence. We also thank Fernando Marchesano and Alessio Notari for useful comments. GB thanks Raphael Flauger and Alexander Westphal for early discussions  about PBH production in axion monodromy inflation during the 2017 Benasque workshop {\it ``Understanding cosmological observations''}. The work of GB and JR is funded by a {\it Contrato de Atracci\'on de Talento (Modalidad 1) de la Comunidad de Madrid} (Spain), with number 2017-T1/TIC-5520, by {\it MINECO} (Spain) under contract FPA2016-78022-P, {\it MCIU} (Spain) through contract PGC2018-096646-A-I00 and by the IFT UAM-CSIC Centro de Excelencia Severo Ochoa SEV-2016-0597 grant. GB thanks the CERN Theory Department and the Leinweber Center for Theoretical Physics of the University of Michigan for hospitality.

\appendix

\section{Cosmological parameters} \label{cosmoparams} 

We use cosmological parameters from the latest (2018) Planck collaboration analysis. For the sake  of completeness, we choose the values from the data set labeled ``TT,TE,EE+lowE+lensing+BAO'' in \cite{Aghanim:2018eyx}, which include (non-CMB) measurements of the baryon acoustic oscillation (BAO) scale. All the numbers in this appendix are given at $68\%$ c.l.\ and correspond to $k=0.05$~Mpc$^{-1}$ (whenever applicable) unless it is specified otherwise. Assuming the base $\Lambda$CDM model as defined by the Planck collaboration, the cosmological parameters we need are \cite{Aghanim:2018eyx}:
\begin{align}
z_{eq} & =3387 \pm 21\,,\\
\Omega_m^0 & =0.3111\pm 0.0056\,,\\ 
\Omega_{\rm DM}^0h^2 & = 0.120\pm0.001 \\
H_0  & =67.66 \pm 0.42\,,\\
n_s & = 0.9665 \pm 0.0038\,, \\
10^9\,A_s & = 2.105 \pm 0.030\,.
\end{align}
If we assume that all neutrinos are non-relativistic today (see the discussion in \cite{Kalaja:2019uju}), then we can use the CMB temperature from FIRAS data  \cite{2009ApJ...707..916F}: $T_{\rm CMB}^0=2.7255\pm0.0006$\,K to obtain the value of the critical density for radiation today, $\Omega_r^0\simeq5.8\cdot10^{-5}$. 

Allowing for a non-negligible running of the scalar spectral index ($dn_s/d\log k \neq 0$) and tensor-to-scalar ratio ($r\neq 0$), and including also Bicep2/Keck Array data \cite{Array:2015xqh}, the constraints on the primordial parameters that are relevant for the global shape of the inflationary potential are, as given in \cite{Akrami:2018odb}:
\begin{align}
r < 0.076\,\, {\rm at}\,\, & k=0.05\,{\rm Mpc}^{-1}\,\, {\rm and}\,\, 95\%\,\, {\rm c.l.}\,,\\
r  < 0.072\,\, {\rm at}\,\, & k=0.002\,{\rm Mpc}^{-1}\,\, {\rm and}\,\, 95\%\,\, {\rm c.l.}\,,\\
n_s & =0.9658 \pm 0.0038\,, \\
d n_s/ d\log k & = -0.0065 \pm 0.0066\,.
\end{align}
The central value and error on $n_s$ are quite robust under the addition of these extra parameters to the base $\Lambda$CDM model.  CMB data alone lead to a somewhat a lower central value for $n_s$ and less constraining upper bounds on $r$. Indeed, removing ``BAO'' data the Planck collaboration obtains: 
\begin{align}
r < 0.079\,\, {\rm at}\,\, & k=0.05\,{\rm Mpc}^{-1}\,\, {\rm and}\,\, 95\%\,\, {\rm c.l.}\,,\\
r  < 0.077\,\, {\rm at}\,\, & k=0.002\,{\rm Mpc}^{-1}\,\, {\rm and}\,\, 95\%\,\, {\rm c.l.}\,,\\
n_s & =0.9640 \pm 0.0043\,, \\
d n_s/ d\log k &= -0.0071 \pm 0.0068\,.
\end{align}

\bibliography{biblio.bib}
\bibliographystyle{hunsrt}

\end{document}